\documentclass[a4paper,11pt
]{article}

\usepackage{jheppub} 

\usepackage{graphicx}

\usepackage{amsmath}
\usepackage[utf8]{inputenc}

\usepackage{amssymb}
\usepackage{amsfonts}
\usepackage{amsbsy}
\usepackage{url}
\usepackage{enumerate}
\usepackage{caption}
\usepackage{subcaption}
\usepackage{color}
\usepackage{ulem}
\usepackage{bm}
\usepackage{comment}
\usepackage{multirow,bigdelim}

\newcommand{\be}{\begin{eqnarray}}
\newcommand{\ee}{\end{eqnarray}}
\newcommand{\p}{\partial}

\makeatletter
\@addtoreset{equation}{section}
\makeatother

\newcommand{\bee}{\begin{equation}}
\newcommand{\eee}{(\end{equation})}

\newcommand{\Tr}{\mathrm{Tr}}

\newcommand\rsout{\bgroup\markoverwith{\textcolor{red}{\rule[0.5ex]{2pt}{0.4pt}}}\ULon}

\newcommand{\acal}{\mathcal{A}}
\newcommand{\lcal}{\mathcal{L}}

\newcommand{\wcal}{\mathcal{W}}
\newcommand{\ycal}{\mathcal{Y}}

\newcommand{\gcal}{\mathcal{G}}

\newcommand{\bcal}{\mathcal{B}}

\newcommand{\xcal}{\mathcal{X}}

\newcommand{\vcal}{\mathcal{V}}
\newcommand{\mcal}{\mathcal{M}}

\newcommand{\diag}{\mathrm{diag}}

\title{SM gauge fields
localized on non-Abelian vortices 
in 6 dimensions
}

\author[a]{Masato Arai}
\author[bc]{, Filip Blaschke}
\author[de]{, Minoru Eto}
\author[d]{, Masaki Kawaguchi}
\author[e]{, Norisuke Sakai}

\affiliation[a]{Faculty of Science, Yamagata University, 
Kojirakawa-machi 1-4-12, Yamagata, 
Yamagata 990-8560, Japan}

\affiliation[b]{Research Centre for Theoretical Physics and Astrophysics, Institute of Physics, Silesian University in Opava, Bezru\v{c}ovo n\'am. 1150/13, CZ-746~01 Opava, Czech Republic}

\affiliation[c]{Institute of Experimental and Applied Physics, Czech Technical University in Prague, Husova 240/5, 110 00 Prague 1, Czech Republic}

\affiliation[d]{Department of Physics, Yamagata University, 
Kojirakawa-machi 1-4-12, Yamagata,
Yamagata 990-8560, Japan}
\affiliation[e]{Department of Physics, and Research and 
Education Center for Natural Sciences, 
Keio University, 4-1-1 Hiyoshi, Yokohama, Kanagawa 223-8521, Japan}

\emailAdd{arai(at)sci.kj.yamagata-u.ac.jp, filip.blaschke(at)physics.slu.cz,
meto(at)sci.kj.yamagata-u.ac.jp, ddwbb.daigaku(at)gmail.com, norisuke.sakai(at)gmail.com
}

\abstract{
A brane-world $SU(5)$ GUT model with global non-Abelian vortices is 
constructed in six-dimensional spacetime. 
We find a solution with a vortex associated to $SU(3)$ separated 
from another vortex associated to $SU(2)$. 
This $3-2$ split configuration achieves a geometric Higgs mechanism 
for $SU(5)\to SU(3)\times SU(2)\times U(1)$ symmetry breaking. 
A simple deformation potential induces a domain wall between 
non-Abelian vortices, leading to a linear confining potential. 
The confinement stabilizes the vortex separation moduli, and 
assures the vorticity of $SU(3)$ group and of $SU(2)$ group to 
be identical. 
This dictates the equality of the numbers of fermion zero modes 
in the fundamental representation of $SU(3)$ (quarks) and of 
$SU(2)$ (leptons), leading to quark-lepton generations. 
The standard model massless gauge fields are localized on the 
non-Abelian vortices thanks to a field-dependent gauge kinetic 
function. 
We perform fluctuation analysis with an appropriate gauge fixing 
and obtain a four-dimensional effective Lagrangian of unbroken 
and broken gauge fields at quadratic order. 
We find that $SU(3) \times SU(2) \times U(1)$ gauge fields are 
localized on the vortices and exactly massless. 
Complications in analyzing the spectra of gauge fields with 
the nontrivial gauge kinetic function are neatly worked out 
by a vector-analysis like method. 
}
\preprint{YGHP-21-2}

\begin{document}
\maketitle


\section{Introduction}
\label{sec:intro}

One of the most interesting paradigms of unified theories beyond 
the standard model is the brane-world scenario in which we live 
on a brane in higher dimensional spacetime 
\cite{ArkaniHamed:1998rs,Antoniadis:1998ig,Randall:1999ee}. 
As candidates for the brane, topological solitons have a number 
of attractive features in contrast with a thin delta-function like 
branes, which may be regarded as somewhat artificial idealizations. 
Traditional reasons repeatedly emphasized in the literature are : 
\begin{enumerate}
\item
The topological solitons are dynamically generated as a consequence 
of spontaneously broken symmetries. 
This phenomenon is referred to as dynamical compactification 
\cite{Dvali:1996bg}. 
\item 
They trap chiral fermions on their world volumes which are topological 
states and therefore inevitably appear irrespective of any details 
\cite{Jackiw:1975fn,Rubakov:1983bb}. 
\end{enumerate}
The models using the topological solitons are often called fat 
brane-world models, and enjoy an additional merit, 
which is common to brane-world scenario. 
The hierarchy problem of fermion masses can be naturally explained 
as a consequence of overlapping of wave functions of the localized 
fermions and the Higgs field \cite{ArkaniHamed:1999dc}.

In order to realize the standard model (SM) on a brane, a serious obstacle has been localization of
massless gauge fields on the brane.
Suppose the gauge symmetry $H$ is broken in the bulk (Higgs phase) and is restored only in the vicinity of the topological soliton.
One might expect that massless $H$ gauge bosons will appear inside the soliton. However, this is not true,
and in general, they get masses of the order of the inverse width of the topological solitons \cite{Dvali:1996xe}.
This is because the bulk except for the soliton core is a sort of non-Abelian superconductor.
Therefore, though the gauge symmetry is restored in the soliton, electric fluxes from a probe electric 
charge put inside the soliton are immediately absorbed 
into the superconducting bulk. Hence, the gauge fields can only propagate for a distance about the width of the soliton, 
namely they are massive~\cite{Dvali:1996xe,Antoniadis:1998ig}.
This argument is quite reasonable, and therefore localizing massless gauge bosons on the soliton generally seem to be quite difficult.

The key idea getting over the difficulty was first proposed by Ref.~\cite{Dvali:1996xe}. 
It employes a sort of electromagnetic duality.
If we put a probe magnetic charge inside the soliton in the superconducting bulk,
the magnetic flux is entirely repelled from the bulk by the Meissner effect, and because of
conservation of the flux, the field lines extend to infinity along the soliton \cite{Dvali:1996xe,Antoniadis:1998ig}.
In Ref.~\cite{Dvali:1996xe} a dual picture of this, namely replacing the Higgs phase by
a confining phase under the assumption that magnetic charges are condensed in the bulk, was proposed;
A massless $H$ gauge field is localized
inside the soliton if the $H$ gauge group is unbroken inside the soliton and is enhanced to a large non-Abelian group $G$
in the confining bulk.
This mechanism, the so-called Dvali-Shifman (DS) mechanism, was rigorously proven in the superYang-Mills theory 
in four dimensions in Ref.~\cite{Dvali:1996xe}, but the higher dimensional version has not yet successfully been proven 
because we do not know how confinement works in higher dimensions. 
Then, a lot of papers followed \cite{Dvali:1996xe} and investigated fat brane scenarios but most of them needed to 
assume the DS mechanism to work.

Ref.~\cite{Antoniadis:1998ig} is one of the earliest works which proposes a simple phenomenological model for the DS mechanism.
There, the Abrikosov-Nilsen-Olsen vortex is considered as a fat 3-brane in six dimensions.
In addition to the SM fields, the model includes an extra scalar field $T$ which condenses 
only inside the vortex core. 
Furthermore, the model has a phenomenological dress factor to the gauge kinetic term
of the form $\Lambda^{-2}\,{\rm Tr}\,T^2 G_{\mu\nu}G^{\mu\nu}$ where $\Lambda$ is a cut-off scale.
Then it was assumed that the model meets the following three conditions:
1) Outside the vortex the SM gauge group $H$ is extended into a larger non-Abelian gauge group $G$
(Note that $H$ is not broken everywhere, and it is included as a part of the unbroken large group $G$ in the bulk).
2) There is no light matter in the bulk.
3) The tree-level gauge coupling (corresponding to the factor $T^2$ in the above example) 
becomes large away from the vortex core.
Ref.~\cite{Antoniadis:1998ig} proposed that when these conditions are satisfied 
the localization of massless $H$ gauge fields on the vortex takes place.

We should note that,
although the seminal work of Ref.~\cite{Dvali:1996xe} gave the basic idea for the localization of massless gauge fields
on fat branes, the detailed analysis on how to get the physical mass spectrum on vortices in six dimensions 
has not yet been given in the literature.
The purpose of this paper is to provide concrete phenomenological 
models for the fat brane-world scenario using topological vortices in six
dimensions along the line of Ref.~\cite{Antoniadis:1998ig} 
and to give a systematic analysis on the physical mass spectra including not only massless but also
massive modes.

A classical realization of a confining vacuum in the bulk can be given 
in terms of generic nonlinear kinetic term, namely a field-dependent gauge kinetic function \cite{Ohta:2010fu,Arai:2012cx,Arai:2013mwa,
Arai:2017lfv,Arai:2017ntb,Arai:2018rwf,Arai:2018uoy,Eto:2019weg} of the form
\be
- \frac{\bcal^2}{4}F_{MN}F^{MN},
\label{eq:fdgkt}
\ee
where $M,N$ are spacetime indices.
If $\bcal$ is a constant, then this is a usual minimal gauge 
kinetic term with a coupling constant $1/\bcal$, but we allow 
$\bcal$ to be a function of scalar fields, such as $\bcal(T) = T$ in the above example.
The scalar fields are not necessarily constant in the extra 
dimensions, but can be a non-trivial function of coordinates 
of extra dimensions as a consequence of dynamics of the system 
under consideration. 
If this is the case, the inverse effective gauge coupling $\bcal$ 
is no longer constant and depends on extra-dimensional 
coordinates nontrivially. 
With a series of our previous works \cite{Arai:2012cx,Arai:2013mwa,
Arai:2017lfv,Arai:2017ntb,Arai:2018rwf,Arai:2018uoy,Eto:2019weg}, we have 
established a general criterion to obtain massless gauge fields 
on the topological solitons: if $\bcal^2$ is square integrable 
with respect to the integration over the whole extra dimensions, 
the massless four-dimensional gauge fields are localized on the 
topological soliton. 
This corresponds to the third criterion 3) of Ref.~\cite{Antoniadis:1998ig} mentioned above.
Since the square-integrability does not depend on details 
of the model, this localization mechanism is robust. 
We verified this statement in various concrete models in five 
dimensional spacetime \cite{Arai:2012cx,Arai:2013mwa,
Arai:2017lfv,Arai:2017ntb,Arai:2018uoy}, and also gave a formal 
proof in generic spacetime dimensions \cite{Arai:2018rwf,Eto:2019weg}. 
With the nontrivial kinetic function, we also found an interesting 
localization mechanism of massless scalar fields \cite{Arai:2018hao} 
on domain walls similarly to Eq.~(\ref{eq:fdgkt}) for gauge fields. 
The mechanism is found to have a topological nature similarly 
to the Jackiw-Rebbi's topological mechanism \cite{Jackiw:1975fn} 
for localization of massless fermions on domain walls.
The fat-brane scenario has been much discussed mostly with 
other contexts, and has produced various different models with 
their own advantages/disadvantages \cite{Oda:2000zc, Dvali:2000rx, 
Kehagias:2000au, Oda:2000dd, Oda:2001ux, Oda:2001yx, Dubovsky:2001pe, 
Ghoroku:2001zu,Akhmedov:2001ny, Kogan:2001wp, Abe:2002rj, 
Laine:2002rh, Batell:2006dp, Guerrero:2009ac, 
Cruz:2010zz, Chumbes:2011zt, Germani:2011cv, Delsate:2011aa, 
Cruz:2012kd, Herrera-Aguilar:2014oua, Zhao:2014gka, Vaquera-Araujo:2014tia,
Alencar:2014moa}.

Since models in five dimensional spacetime is simplest to analyze, 
there have been many concrete brane-world models using domain walls 
to obtain the symmetry breaking of grand unified theories (GUT) 
down to the SM, such as in \cite{Davies:2007xr,
Davidson:2007cf,Thompson:2009uk,Callen:2010mx,Callen:2012kd,
Okada:2017omx,Okada:2019fgm}, 
without an explicit mechanism to localize the gauge fields. 
With our mechanism for localization of gauge fields, we have 
constructed a concrete model for the $SU(5)$ GUT on domain 
walls \cite{Arai:2017lfv}. 
The GUT symmetry breaking is determined by positions of domain 
walls. This geometric Higgs mechanism is a characteristic feature 
of the brane-world model with topological solitons. 
By introducing a moduli stabilization potential we obtained 
the $SU(5)\to SU(3)\times SU(2)\times U(1)$ leading to the SM. 
As an alternative possibility, we have also obtained a five-dimensional 
model for the SM, where the condensation of the SM Higgs field $\Phi$ is driven 
by the formation of domain walls which localizes the SM. 
It predicts a new contribution to the $\Phi\to \gamma \gamma$ 
decay and the possible experimental accessibility of heavy monopoles 
\cite{Arai:2018uoy}. 

To explain quark lepton generations naturally, however, we need 
an additional idea in the brane-world models with topological 
solitons. 
One such mechanism is to use vortices in the extra dimensions. 
If there is a vortex with $k$ vorticity ($k$ coincident vortices), 
the index theorem gives $k$ zero modes for fermions.  
This mechanism has been proposed previously to explain fermion 
generations in the brane-world scenario \cite{Libanov:2005mv,Frere:2000dc}, 
although the model was without the localization mechanism for gauge fields.

The purpose of this paper is to propose a class of models in 
six-dimensional spacetime which is a non-Abelian 
generalization of the simple model considered in Ref.~\cite{Antoniadis:1998ig}.
Our model is based on a GUT-inspired $G=SU(5)$ gauge theory with 
a field-dependent gauge kinetic function in six spacetime dimensions. 
It contains complex scalar fields in a singlet and 
an adjoint representation of $SU(5)$, which are combined into 
a $5\times 5$ matrix-valued field $T$. 
The model admits topologically stable non-Abelian global vortices.
We find that the massless gauge fields of
$H = SU(3)\times SU(2) \times U(1)$ are localized on the 
four-dimensional world volume of the non-Abelian global vortices 
with multiple winding numbers.\footnote{
An application of the non-Abelian {\it local} vortices to the 
brane-world physics without localization of the gauge fields was 
considered in Ref.~\cite{Eto:2004ii}.}

By a simple potential with the $5\times 5$ matrix field $T$ 
we can obtain vortex equations for each diagonal components 
$T = \diag(T_1,T_2,T_3,T_4,T_5)$ independent of each other. 
Then we obtain five different species of vortices $I=1,\cdots, 5$ 
corresponding to vortices in the $I$-th diagonal component $T_{I}$. 
We find two important phenomena due to 
the vortices: 
\begin{enumerate}
\item Geometric Higgs mechanism 

The gauge symmetry $G=SU(5)$ is partially broken when the topological vortices are present. 
Interestingly, the breaking pattern of the $SU(5)$ gauge symmetry depends on positions 
of vortices.
Namely, dynamics of the non-Abelian vortices determines
which subgroup of $SU(5)$ remains unbroken. 
We are primarily interested in the configuration where vortices 
in the three diagonal components, say $T_{1}, T_{2}, T_{3}$, 
of the matrix field $T$ are coincident at a point in the extra 
dimensional plane, whereas vortices in 
the remaining two diagonal components $T_{4}, T_{5}$ are 
coincident at another point. 
This provides a symmetry breaking pattern 
$SU(5) \to SU(3) \times SU(2) \times U(1)$. 
It is important to notice that the origin of the symmetry breaking 
resides only locally near the non-Abelian vortices. 
This is the reason why the SM gauge fields are localized around 
the vortices. 
We call this vortex solution as the $3-2$ splitting configuration. 
Note that the first criterion 1) of Ref.~\cite{Antoniadis:1998ig} mentioned above is
naturally satisfied by the generation of the non-Abelian vortices.
\item Confinement of non-Abelian global vortices 

The separation between the position of one group of vortices 
in $T_{1}, T_{2}, T_{3}$ and position of another group in 
$T_{4}, T_{5}$ is a moduli that depends on the 
details of the scalar potential, namely on a 
particular ratio of potentials for singlet and adjoint in parts of $T$. 
If we perturb this ratio, 
we find that separation is no longer a moduli. 
Namely a potential energy is induced and becomes 
constant for asymptotically large separations. 
Thus a confining force emerges between non-Abelian global vortices. 
The confining force can only end at another non-Abelian global 
vortex. This process continues until they finally form an $SU(5)$ 
singlet combination of non-Abelian vortices. 
Namely only when each diagonal component has the same 
number of vortices, the configuration becomes stable. 
Any vortices which cannot form a singlet combination are 
dynamically removed to spatial infinity by the confining energy 
density (a domain wall) extending to infinity. 
This is important once fermions couple to the vortices, because 
the number of fermion zero modes is identical to the winding number. 
Hence, the confinement of non-Abelian global vortices ensures 
that the number of fermion zero modes is common for different 
representations of the SM, namely leptons and quarks should come 
in generations. 
\end{enumerate}

We also obtain a low energy effective theory on the background 
of the $3-2$ splitting configuration with a particular focus on 
the problem of localization of the unbroken 
$SU(3)\times SU(2) \times U(1)$ gauge fields.
Deriving the effective Lagrangian can be quite complicated 
even at the quadratic order of small fluctuations, because 
of a number of reasons. 
\begin{enumerate}
\item
 The background solution is non-trivial, namely non-Abelian vortices, 
\item
The gauge kinetic Lagrangian in Eq.~(\ref{eq:fdgkt}) is not in 
the canonical form.  
\item
Apart from $SU(3) \times SU(2) \times U(1)$ four-dimensional vector fields, 
we have gauge fields corresponding to the broken generators of $SU(5)$. 
The extra-dimensional components of gauge fields are also present, 
and mix among themselves and with the matrix-valued complex scalar fields. 
\item
We have to take care of these issues by choosing an appropriate gauge fixing.  
\end{enumerate}
In order to organize the calculations, 
we develop an effective and a compact formula generalizing the 
usual three-dimensional vector analysis.
It turns out that this approach is useful to clean up complicated calculations 
and make things transparent.
Furthermore, our method allows us to treat both unbroken and 
broken parts in a very similar manner.
Armed with our vector-analysis-like method, we show most importantly 
that the 3-2 splitting configuration 
of non-Abelian global vortices localizes the massless degrees of 
freedom corresponding to the $SU(3) \times SU(2) \times U(1)$ SM 
gauge fields on the four-dimensional world-volume of the vortices. 
We also show that other fields, except for a mixing part of 
the extra-dimensional component of the broken gauge fields and 
the scalar fields, are either massive or unphysical 
in the sense that they are absorbed by massive Kaluza-Klein (KK) 
towers of the gauge fields.
Hence, our 3-2 splitting background configuration of the 
non-Abelian global vortices in six dimensions is a promising 
platform for a fat brane-world scenario with GUT.

This paper is organized as follows. 
In Sec.~\ref{sec:simplest_non-Abelian_vortex}, we present our model 
admitting non-Abelian global vortices in six dimensions, and 
study multiple vortices, especially the 3-2 splitting configuration 
that leads to the desired symmetry breaking 
$SU(5) \to SU(3) \times SU(2) \times U(1)$. 
In Sec.~\ref{sec:improve}, the confinement of non-Abelian global 
vortices by domain walls are described. 
Sec.~\ref{sec:fluctation} is devoted to derive a four-dimensional 
low energy effective Lagrangian of the gauge fields. 
We develop the useful vector-analysis-like technique and examine 
which fields provide massless/massive, physical/unphysical, 
and normalizable/non-normalizable modes under the 3-2 splitting 
background.
We give proofs of some theorems resembling the Helmholtz's 
theorem for our vector-analysis-like method 
in Appendix~\ref{sec:app1}. 


\section{
A brane-world model with non-Abelian vortices}
\label{sec:model}
\subsection{
Non-Abelian global vortices for $SU(5)\to SU(3)\times SU(2)\times U(1)$ }
\label{sec:simplest_non-Abelian_vortex}

Let us consider an $SU(5)$ non-Abelian gauge theory in 6 dimensions 
\be
{\cal L} = \Tr\left[- \frac{{\cal B}(T){\cal B}^\dagger(T)}{2} 
{\cal F}_{MN}{\cal F}^{MN} + D_MT (D^MT)^\dagger\right] - V ,
\label{eq:lag}
\ee
where $M=0, 1, \cdots, 5$ denotes spacetime indices, 
${\cal F}_{MN} = \p_M {\cal A}_N - \p_N {\cal A}_M 
+ i \left[{\cal A}_M,{\cal A}_N\right]$ 
is a field strength of $SU(5)$ gauge fields ${\cal A}_M$.
A $5\times 5$ matrix-valued complex scalar field $T$ contains 
a singlet $T_0$ and an adjoint $\hat T$ representation of $SU(5)$ 
\be
T_0 = \Tr\, T,\quad \hat T = T - \frac{T_0}{5} {\bf 1}_5, 
\ee
with $\Tr~\hat T = 0$.
Then a covariant derivative is defined by
\be
D_M T = \p_M T + i \left[{\cal A}_M,T\right].
\ee

The Lagrangian (\ref{eq:lag}) has a peculiar factor 
${\cal B}\bcal^\dag$ in front of the gauge kinetic term ${\cal F}_{MN}^2$.
If we take a constant ${\cal B} = 1/g$, the Lagrangian is just 
standard one. 
Instead, we allow ${\cal B}$ to be a generic function of $T$, 
and call it a gauge kinetic function. 
We require ${\cal B}$ to be at least a Hermitian 
matrix and invariant under the $SU(5)$ gauge transformation. 
Otherwise, we leave it arbitrary for now, 
since it does not play 
any role in the rest of this section.\footnote{
One can jump to Eq.~(\ref{eq:emp_B}) to see a concrete example for $\bcal$.}
It will play an important role in the next section, and 
we will clarify concrete conditions on ${\cal B}$ for a physically 
meaningful brane-world GUT model, namely the conditions for having
massless gauge bosons localized on non-Abelian vortices.\footnote{
The conditions for $\bcal$ are given in Eqs.~(\ref{eq_T_bg}) and (\ref{eq:gauge-coupling}).}
In the next section we will conclude that $\bcal\bcal^\dagger$ has to asymptotically go to zero far away from the solitons,
and it implies that the effective gauge coupling $g \sim 1/\sqrt{\bcal\bcal^\dagger}$ diverges.
Therefore, the Lagrangian (\ref{eq:lag}) is not suitable for describing physics in a homogeneous vacuum.
Instead, we interpret it as a phenomenological model which 
is suitable to describe non-trivial confinement physics (the DS mechanism) under the presence of topological solitons
in higher dimensions, as
proposed in Ref.~\cite{Antoniadis:1998ig}.

There also seems to be no a priori condition for the scalar 
potential $V$ of $T$ except for the gauge invariance and reality 
condition.
Hence, it can be an arbitrary function of gauge invariant quantities 
such as $|T_0|^2$, $\Tr\left[\hat T\hat T^\dagger\right]$, 
$\Tr\left[(\hat T\hat T^\dagger)^2\right]$, 
$T_0^*\,\Tr\left[\hat T^2\hat T^\dagger\right] + {\rm h.c.}$, 
$T_0^{*2}\,\Tr\left[\hat T^2\right]  + {\rm h.c.}$ and so on.
Instead of surveying such vast possibilities, we concentrate 
on a simple potential and its deformations to obtain a platform 
for the brane-world and GUT 
\be
V = \frac{\lambda^2}{2}\Tr\left[\left(TT^\dagger 
- v^2 {\bf 1}_5\right)^2\right].
\label{eq:pot0}
\ee
The vacuum configuration of $V$ up to symmetry transformation 
is obviously 
\be
T = v{\bf 1}_5.
\ee
This is the $SU(5)$ preserving vacuum, but it is important to 
realize that the $U(1)$ global symmetry $T \to e^{i\alpha}T$ 
is spontaneously broken. 
Hence, the vacuum manifold is isomorphic to $S^1$ which is not 
simply connected space, and the fundamental homotopy group is 
nontrivial as $\pi_1(S^1) = \mathbb{Z}$. 
This gives rise to global vortices (three-branes in six-dimensional 
spacetime) which are topologically stable. 
The Euler-Lagrange equations can be solved by consistently setting 
all the off-diagonal components of $T$ and gauge fields to vanish. 
This leads to five decoupled Euler-Lagrange equations for 
the diagonal elements of $T = \diag(T_1,T_2,T_3,T_4,T_5)$. 
Assuming $T_I$ depends only on extra-dimensional coordinates 
$x^a, a=4, 5$, we obtain 
\be
\partial_a^2 T_I - \lambda^2\left(|T_I|^2 -v^2\right)T_I = 0, 
\quad I=1, \cdots, 5
\label{eq:eom_global_vortex}
\ee
Each equation is identical to the familiar equation for a global 
vortex, which can be derived from a one scalar model 
\be
{\cal L}' = \left|\partial_M \phi\right|^2 
- \frac{\lambda^2}{2}\left(|\phi|^2 - v^2\right)^2,
\ee
if we identify $\phi$ with $T_I$.

To obtain the $k_I \in \mathbb{Z}$ coincident vortices at the 
origin for the $I$-th diagonal field $T_I$, we can solve the 
angular part of the equation by expressing $T_I$ as 
\be
T_I = vf_I(r) e^{ik_I\theta}, 
\ee
in terms of the polar coordinates $x_4 + i x_5 = r e^{i\theta}$. 
The remaining radial equation is given as 
\be
f_I'' + \frac{f_I'}{r} - \frac{k_I^2}{r^2}f_I- 
\lambda^2v^2(f_I^2-1)f_I = 0.
\ee
This should be solved under the following boundary condition 
\be
f_I(0) = 0,\quad f_I(\infty) = 1.
\ee
Although these vortex solutions are mere five copies of the 
standard global vortex solutions embedded into the diagonal 
components of the matrix field $T$, they actually have a  
characteristic property of non-Abelian vortices. 
For example, take the simplest case with $k_1 = 1$ and all the 
others zeros ($k_{2,3,4,5}=0$). 
In terms of the matrix field $T$, the vortex solution is 
\be
T = v\, \diag (f_1e^{i\theta}, 1,1,1,1) 
= e^{i\frac{\theta}{5}} v\, \diag\left(f_1e^{i\frac{4}{5}\theta},
e^{-i\frac{1}{5}\theta},e^{-i\frac{1}{5}\theta},
e^{-i\frac{1}{5}\theta},e^{-i\frac{1}{5}\theta}\right).
\label{eq:10000}
\ee
At spatial infinity ($f_1 \to 1$) we have 
\be
T\big|_{r\to \infty} = v e^{i\frac{\theta}{5}} 
e^{i\frac{4\theta}{5}\lambda_{14}},
\ee 
with an $SU(5)$ generator $\lambda_{14} = \diag\left(1,-\frac{1}{4},
-\frac{1}{4},-\frac{1}{4},-\frac{1}{4}\right)$.
This decomposition shows that the winding number in the overall 
$U(1)$ group is $\frac{1}{5}$ when we go around the vortex once.
In this sense, the vortex is called $\frac{1}{5}$ fractionally quantized 
global vortex. Let us turn our eyes on the opposite limit $r=0$. 
We have 
\be
T\big|_{r\to 0} = v\, \diag (0, 1,1,1,1) .
\ee
Namely symmetry breaking $SU(5) \to SU(4) \times U(1)$ occurs 
at the very center of the vortex. 
This implies that the vortices transforms nontrivially 
under the non-Abelian global $SU(5)$ transformations, and 
such vortices are called non-Abelian vortices. 
In summary, the vortex solution in this model is the $\frac{1}{5}$ 
fractionally quantized non-Abelian global vortex.
Note that the $SU(5)$ gauge fields do not play any role at all, 
and therefore the above vortex solutions are called 
non-Abelian global vortices as found in models without gauge 
fields \cite{Balachandran:2002je,Nitta:2007dp,
Nakano:2007dq,Eto:2009wu,Eto:2013bxa}.

Since the equations in Eq.~(\ref{eq:eom_global_vortex}) for five 
diagonal fields $T_I$ are decoupled, we can freely choose positions 
of the vortices in each diagonal component $T_I$. 
Namely, they are moduli of the solutions. 
Let us consider a solution where the first three diagonal components, 
say $T_1, T_2, T_3$, have a common number of vortices at a single 
position, whereas the remaining two diagonal components ($T_4, T_5$) 
have also the same number of the vortices at another position 
\be
T = v\left(
\begin{array}{cc}
f_3(r_3)e^{ik_3\theta_3} {\bf 1}_3 & 0 \\
0 & f_2(r_2)e^{ik_2\theta_2}{\bf 1}_2
\end{array}
\right),
\label{eq:3-2}
\ee
where $(r_3,\theta_3)$ and $(r_2,\theta_2)$ are the polar coordinates 
whose origins are at the respective vortex centers. 
We have $T \to v\,\diag(0,0,0,e^{i\theta_2},e^{i\theta_2})$ at $r_3 \to 0$,
and $T \to v\,\diag(e^{i\theta_3},e^{i\theta_3},e^{i\theta_3},0,0)$ 
at $r_2 \to 0$. 
Since vortex positions of $T_1, T_2, T_3$ and that of $T_4, T_5$ are 
distinct, the solution breaks $SU(5)$ down to 
$SU(3)\times SU(2) \times U(1)$. 
We can identify this gauge symmetry breaking as the breaking of GUT 
to the SM. 
We should emphasize that this symmetry breaking occurs only 
locally in the vicinity of the vortex centers. 
This fact is important for having massless gauge fields 
{\it localized} on the vortices as we will see later.

We note that tensions of the non-Abelian global vortices in our model are logarithmically
divergent similarly to usual Abelian global vortices. However, it is not harmful for construction of the
effective theory of four-dimensional fields as we will explain in the subsequent sections.

\subsection{Moduli stabilization through confinement of vortices}
\label{sec:improve}

Though we are satisfied with the vortex solution (\ref{eq:3-2}), of course it is not a perfect solution.
Firstly, the vortex positions are moduli. Therefore, they in general scatter around, and then
breaking pattern of the gauge symmetry becomes $SU(5) \to U(1)^4$ which is not acceptable with respect to the brane world model.
Secondly, the vortex numbers $k_a$ are arbitrary. Indeed, the vortex number is related to the number of generation of 
fermions coupled to the $T$ field. Therefore, we would like to have $k_1 = k_2 = \cdots = k_5$ at least, and, if possible,
we further want $k_a = 3$ for all $a$.
Unfortunately we cannot solve for all these conditions at once, but we can at least solve them partially.

Let us begin with extracting quadratic terms from the potential (\ref{eq:pot0})
\be
V \supset -\lambda^2v^2 \Tr\left[TT^\dagger\right] 
= -\lambda^2v^2\left(\frac{|T_0|^2}{5} + \Tr\left[\hat T\hat T^\dagger\right]\right).
\ee
The particular ratio $\frac{1}{5}$ between two coefficients of $|T_0|^2$ and $\Tr\left[\hat T\hat T^\dagger\right]$ is
important for having the decoupled five equations (\ref{eq:eom_global_vortex}).
However, there is no a priori reason to chose this specific ratio from the symmetry ground.
We can single out, for example, the traceless part and modify the potential adding 
\be
\delta V = \alpha^2\, \Tr\left[\hat T \hat T^\dagger\right].
\label{eq:deltaV}
\ee
To see what happen by this additional term, let us plug the minimal vortex solution (\ref{eq:10000})
with $k_1 = 1$ and $k_{2,3,4,5}=0$ into the additional term.
It reads
\be
\delta V
= \frac{22 v^2\alpha^2}{25}\left(1+f_1^2-2f_1\cos\theta\right) \to \frac{44 v^2\alpha^2}{25}\left(1-\cos\theta\right),
\ee 
as $r\to \infty$. This depends on the angular coordinate $\theta$ via $\cos\theta$. Therefore, when we traverse around
the vortex, we inevitably across the potential barrier once. Namely, the minimal vortex is attached by the domain wall,
as familiar axion cosmic strings. For the domain wall pulls the vortex towards the spatial infinity, we cannot 
retain the vortex. However, the domain wall can end on the different vortex, say, with $k_2=1$ and $k_{1,3,4,5}=0$
at some other point. When we are sufficiently far from both of the vortices, we have $f_1 \sim f_2 \to 1$ and $\theta_1 \sim \theta_2 \equiv \theta$.
Therefore, the additional potential asymptotically reduce to
\be
\delta V 
\to \frac{12v^2\alpha^2}{5}(1-\cos\theta).
\ee 
There is still the domain wall. To eliminate the domain wall completely, we need the same number of the vortices in
all the diagonal components. For example, when $k_{1,2,\cdots,5} = 1$, we have $\hat T \to 0$ and
\be
\delta V \to 0.
\ee
Now, the $\theta$ dependence asymptotically disappears. This can be understood as follows. The domain walls may exist but they are 
completely terminated by the five vortices. In other words, the vortices are confined to form
the singlet (a set of five different vortices) by the linear force of domain wall.\footnote{
This singlet is a sort of baryonic type with $k_I > 0$ or ($k_I < 0$) for all $I$.
We could have a mesonic singlet of $k_1 = 1$ and $k_1 = -1$ with $k_{2,3,4,5}=0$. But it is topologically
trivial and it would annihilate.
} Note that the domain wall appears by adding $|T_0|^2$ instead of $\Tr\left[\hat T\hat T^\dagger\right]$
in Eq.~(\ref{eq:deltaV}). Only when they appear together with a particular ratio so that they are unified as 
$\Tr\left[TT^\dagger\right]$, no domain walls appear even in the single vortex with $k_1=1$ with $k_{2,3,4,5}=0$.
Namely, the vortices are deconfined when the scalar potential can be described by $T$ only (without $T_0$ and/or $\hat T$) as
Eq.~(\ref{eq:pot0}).

The vortex confinement is definitely important piece of solving our problems described at the beginning of this subsection.
Firstly, it provides the attractive confining force among the asymptotically separated vortices. This lifts the position moduli.
Secondly, the domain walls discard unconfined constituent vortices to the spatial infinity. Namely, the vortex winding numbers
$k_1$, $k_2$, $\cdots$, $k_5$ are automatically adjusted to a same integer.
Hence, this ensures unification of generations of the localized fermions.

All is good news so far. However, there are still unclear points.
a) we cannot explain the reason why the unified generation becomes 3.
b) We cannot predict very well what kind of the singlet configuration remains 
under the presence of the confining force. If there is only attractive force
by domain wall, all the vortices would completely coincide. This is not good for
the brane-world scenario since $T \propto {\bf 1}_5$ and $SU(5)$ is never broken. 
We can show this via a concrete numerical simulation.
\begin{figure}[t]
\begin{center}
\begin{tabular}{cc}
\begin{minipage}{0.48\hsize}
\begin{center}
\includegraphics[width=\hsize]{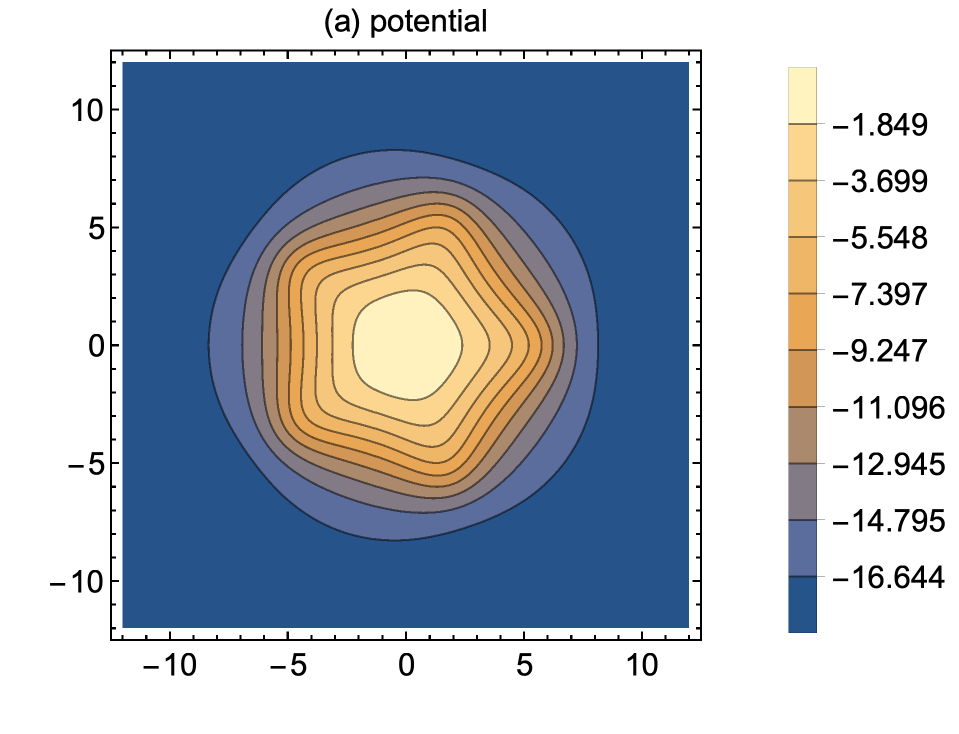}
\end{center}
\end{minipage}
&
\begin{minipage}{0.48\hsize}
\begin{center}
\includegraphics[width=\hsize]{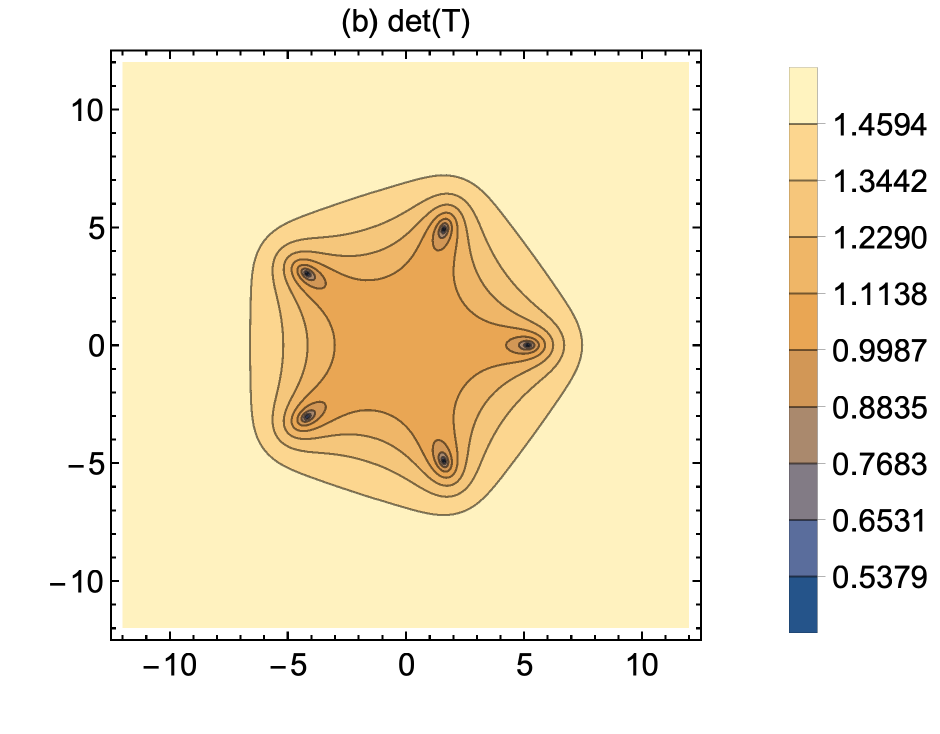}
\end{center}
\end{minipage}\\\ \\\ \\
\begin{minipage}{0.48\hsize}
\begin{center}
\includegraphics[width=\hsize]{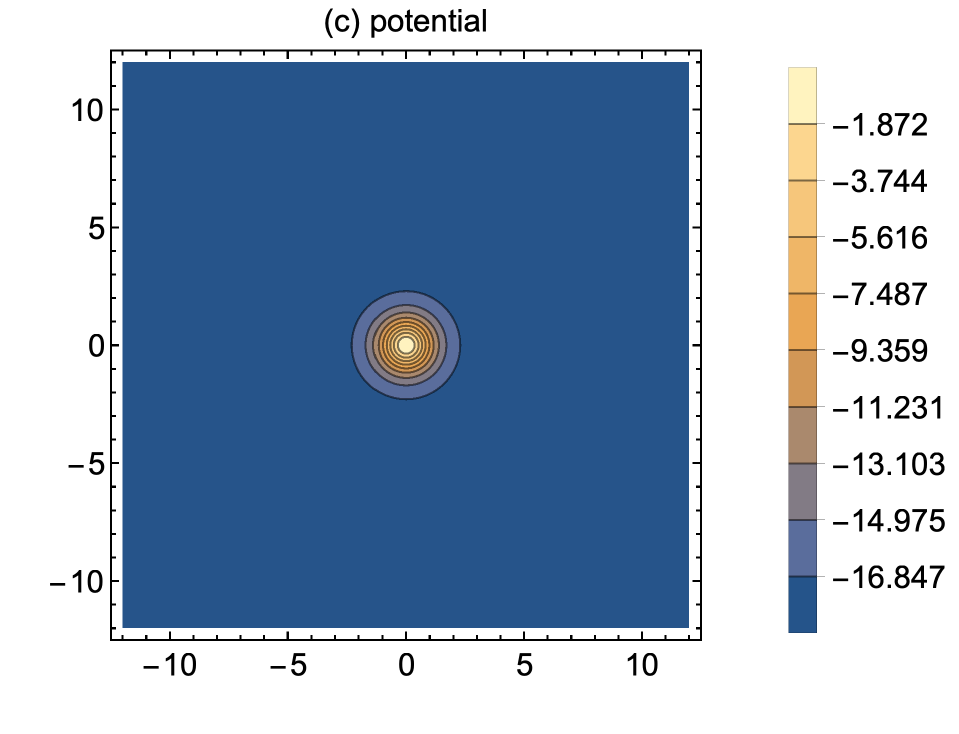}
\end{center}
\end{minipage}
&
\begin{minipage}{0.48\hsize}
\begin{center}
\includegraphics[width=\hsize]{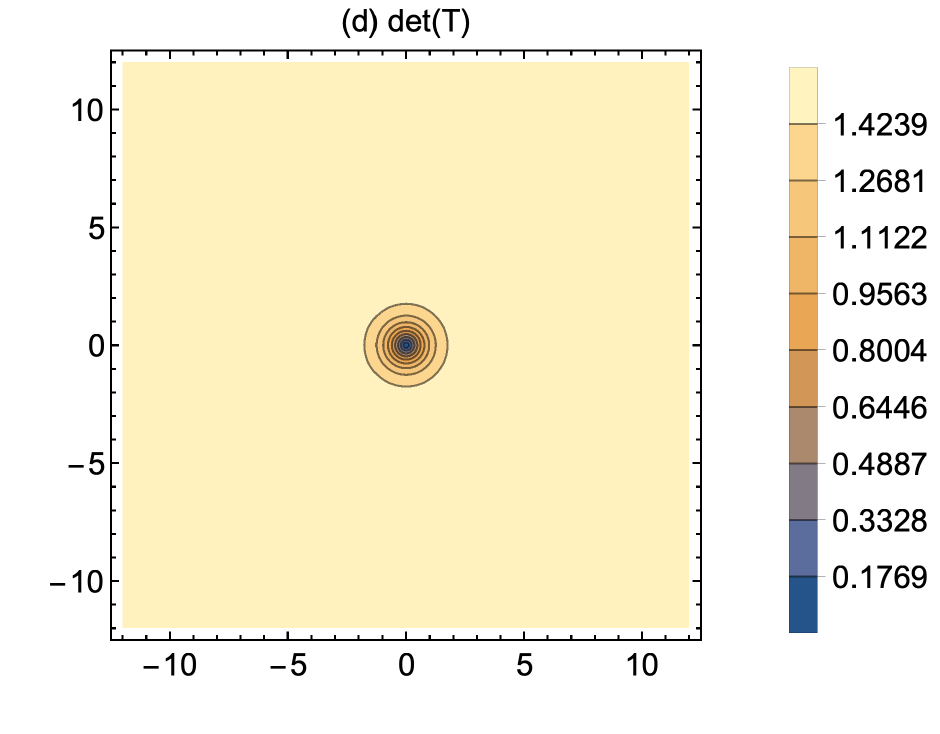}
\end{center}
\end{minipage}
\end{tabular}
\caption{The two snap shots of the relaxation process starting with the five separate vortices. We take $\lambda_0 = 1$,
$v=1$, and $m^2 = 0.3$. The upper two panels are at the early stage and the lower two panels are at the late stage.}
\label{fig:5Vs}
\end{center}
\end{figure}
In Fig.~\ref{fig:5Vs} we show typical configurations including five non-Abelian vortices.
We numerically constructed these by making use of a standard relaxation method [add a dissipation to the EOMs for the potential
$V$ in (\ref{eq:pot0}) with $\delta V$ in (\ref{eq:deltaV})].
We first prepare an appropriate initial configuration with 5 vortices separately placed at vertices of a pentagon.
Then, we run the relaxation simulation. 
We exhibit two snap shots 
at an early [(a) and (b) of Fig.~\ref{fig:5Vs}] and a late [(c) and (d) of Fig.~\ref{fig:5Vs}] stages. 
The panel (b) shows the amplitude $|\det T|$ for well separated five non-Abelian vortices. 
The panel (a) shows the additional potential density 
$\delta V$ which is 
nothing but the domain wall which is in charge of the confinement. 
The panels (c) and (d) show the same informations as (a) and (b) but at much later stage.
The pentagon is completely squashed, and we are left with an integer quantized Abelian vortex.
This occurs because there is only confining force among the non-Abelian vortices.

We should introduce a repulsive force which 
can compensate the domain wall force. A candidate is
\be
\tilde V = \frac{\tilde \lambda^2}{2}\left(\Tr\left[ T T^\dagger- v^2{\bf 1}_5\right]\right)^2
+ \alpha^2\, \Tr\left[\hat T\hat T^\dagger\right].
\label{eq:pot1}
\ee
When $\alpha^2 > 0$, the vacuum is again given by $T = v {\bf 1}_5$.
Note that the first terms looks very similar to Eq.~(\ref{eq:pot0}) but is different.
We can decompose this
\be
\tilde V = 
\frac{\tilde \lambda^2}{2}\left(\frac{1}{5} |T_0|^2 - 5 v^2\right)^2
+\frac{\tilde \lambda^2}{2}\left(\Tr\left[ \hat T \hat T^\dagger\right]\right)^2
+ \tilde \lambda^2\left(\frac{1}{5} |T_0|^2 - 5 v^2 
+ \frac{\alpha^2}{\tilde \lambda^2}\right)\Tr\left[ \hat T \hat T^\dagger\right].
\ee
We should pay attention on the coefficient of $\Tr\left[\hat T \hat T^\dagger\right]$ in the third term.
It is positive ($\alpha^2 > 0$) in the bulk, but becomes negative in the vicinity of the vortex center since
some of the diagonal component of $T$ vanishes and $|T_0|^2$ is smaller than $25 v^2$.
When the sign of the coefficient flip to negative, then $\hat T$ locally condense around the vortices.
This leads to a short range repulsive force.
\begin{figure}[t]
\begin{center}
\includegraphics[width=10cm]{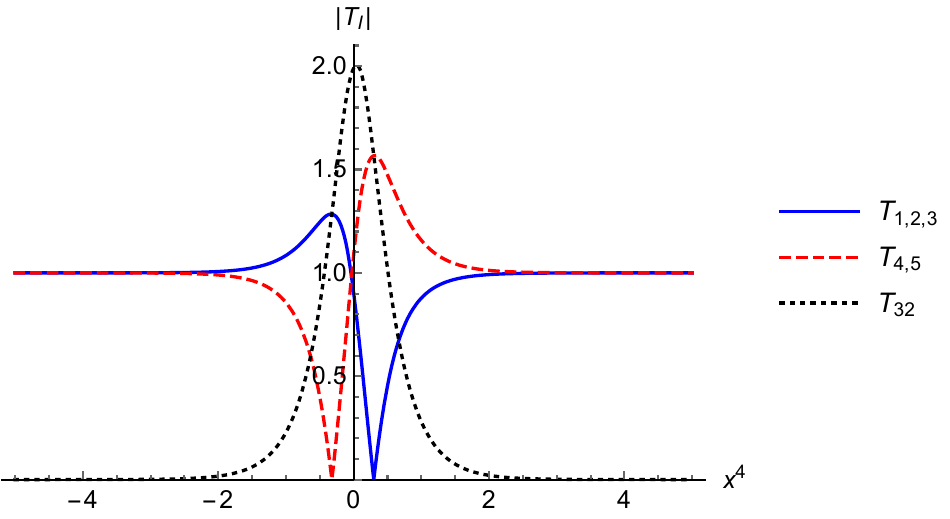}
\caption{A 3-2 splitting of the five vortices. The vortex centers are on the $x^4$ axes and the cross section of the absolute values
$|T_I|$ are shown. The parameters are taken as $\tilde \lambda = 15$, $\alpha = \frac{3}{2}\sqrt{\frac{5}{2}}$, and $v=1$.
The blue curves corresponds to $|T_{1,2,3}|$ and the red dashed lines to $|T_{4,5}|$. The blue dotted curve shows 
$T_{32} = \Tr\left[T \lambda_{32}\right]$.}
\label{fig:3_2V}
\end{center}
\end{figure} 
Fig.~\ref{fig:3_2V} shows a numerical solution where the five vortices split into triply ($T_{1,2,3}$) and doubly $(T_{4,5})$
coincident parts at a finite distance. As can be clearly seen from the figure, the condensations of $T_{4,5}$ increase at
the center of the triply coincident vortices whereas those of $T_{1,2,3}$ also increase near the doubly coincident vortices.
This is a direct consequence of the condensation of $T_{32} \equiv \Tr\left[T\lambda_{32}\right]$ with
$\lambda_{32} = \diag\left(\frac{1}{3},\frac{1}{3},\frac{1}{3},-\frac{1}{2},-\frac{1}{2}\right)$ which is also depicted by the black dotted line in Fig.~\ref{fig:3_2V}.
The blue (red) curves must touch zero at the center of the $T_{1,2,3}$ ($T_{4,5}$) 
vortex but at the same time it tends to increase near the adjacent vortex cores $T_{4,5}$ ($T_{1,2,3}$). 
This opposite tendency conflicts to each other, and generates a desired short range inter-vortex repulsion.
Similar mechanism for having vortex molecules plays 
an important role in a very different context, such as 
in coherently coupled multicomponent BECs of cold atoms 
\cite{Son:2001td,Kasamatsu:2004tvg,Eto:2011wp,Eto:2012rc,
Eto:2013spa,Nitta:2013eaa,Kasamatsu:2015cia,Tylutki:2016mgy,
Eto:2017rfr,Kobayashi:2018ezm,Eto:2019uhe,Yang:2020wes}, 
and also in the dense QCD \cite{Eto:2021nle}.

A remark: Though we found one particular numerical solution with the desired 3-2 splitting structure, we have not performed
any systematic survey on the parameter space. To figure out how common this configuration is is an important future work.
Furthermore, we do not establish stability of our numerical solution. 
We also leave this as another future work. For this paper, we satisfy ourself by the fact that we succeeded in obtaining
the 3-2 splitting solutions in the models with the potentials in Eqs.~(\ref{eq:pot0}) and (\ref{eq:pot1}).

In the next section, we will present a formal analysis on localization of the $SU(5)$ gauge fields.
For that purpose, any concrete solutions are not needed. So we will assume that the 3-2 splitting vortex configuration is
obtained by wisely setting a model up somehow.

\section{Localization of the gauge fields on vortices}
\label{sec:fluctation}


\subsection{Quadratic Lagrangian for gauge fields and gauge-fixing}

In this section we study small fluctuations of the $SU(5)$ gauge 
fields ${\cal A}_M$ in the presence of the non-Abelian global 
vortices, in order to clarify massless and massive modes. 
As mentioned above, we are primarily interested in the 3-2 
splitting background in which the gauge symmetry is spontaneously 
broken as $SU(5) \to SU(3)\times SU(2) \times U(1)$. 
Our starting point is to divide the small fluctuations as 
\be
{\cal A}_{M} = 
\begin{pmatrix}
\gcal_{M} & \xcal_{M}/\sqrt2\\
\xcal_{M}^{\dagger}/\sqrt2 & \wcal_{M}
\end{pmatrix} 
+ \ycal_{M} 
\dfrac{1}{\sqrt{60}} 
\begin{pmatrix}
2 I_{3} & \\
 & - 3 I_{2}
\end{pmatrix}
\label{eq:fluctuation_A}
\ee
where $ \gcal_{M} $ is an $ SU(3) $ gauge field, $ \wcal_{M} $ 
is an $ SU(2) $ gauge field and $ \ycal_{M} $ is a $ U(1) $ gauge field.
The off-diagonal gauge field $ \xcal_{M} $ is a 2 by 3 rectangular 
complex matrix. 
Finding out physical spectra of these gauge fields ${\cal G}_M$, 
${\cal W}_M$,  ${\cal G}_Y$, and  ${\cal X}_M$ 
involves complicated calculation 
due to the following factors:
\begin{enumerate}
\item
Our background is a nontrivial configuration of vortices. 
\item
We need a separate treatment for the gauge fields with the four-dimensional 
indices ${\cal A}_\mu$ and with the extra-dimensional indices 
${\cal A}_a$. 
\item
We also have to distinguish the unbroken gauge fields 
$\{{\cal G}_M, {\cal W}_M, {\cal Y}_M\}$, and the broken gauge 
field ${\cal X}_M$, which absorb the fluctuations from the 
scalar field $T$. 
\item
We have to clarify the distinction between the physical modes 
from the unphysical modes by taking account into gauge invariance. 
\end{enumerate}
Despite these obstacles, we will provide a reasonably simple scheme 
with which we handle complicated fluctuation analysis. 
We can also apply our scheme to a wide class of models, 
since it does not depend on details of the background. 

Our main goal in this section is to examine whether massless gauge 
fields corresponding to the SM gauge group $SU(3)\times SU(2) \times U(1)$ 
are localized on the 3-2 splitting vortex background or not.
The field dependent gauge 
kinetic function ${\cal B}(T)$ in Eq.~(\ref{eq:lag}), which did not play
any role in the previous section, will play a crucial role here.
We assume ${\cal B}$ is a function of $T$ and $T^\dagger$ (more 
generically a function of $T_0$, $T_0^*$, $\hat T$ and $\hat T^\dagger$),
although we do not fix a concrete ${\cal B}$ for now.\footnote{
The constraint on $\bcal$ for having localized massless non-Abelian gauge fields inside the vortices
is given in Eq.~(\ref{eq:gauge-coupling}), and 
a concrete example of $\bcal$ is given in Eq.~(\ref{eq:emp_B}).}
When the background solution $T$ is a diagonal matrix, 
we obtain a diagonal $ 5 \times 5 $ matrix ${\cal B}$: 
\be
T\big|_{\rm bg} = \begin{pmatrix}
\tau_3(x^a) {\bf 1}_3 & \\
& \tau_2(x^a) {\bf 1}_2
\end{pmatrix},
\qquad
{\cal B} \big|_{\rm bg} = \begin{pmatrix}
\beta_3(\tau_3) {\bf 1}_3 & \\
& \beta_2(\tau_2) {\bf 1}_2
\end{pmatrix}
.\label{eq_T_bg}
\ee
For later convenience, let us define 
\be
\beta_1 \equiv \sqrt{\frac{3 |\beta_2|^2 + 2 |\beta_3|^2}{5}},\quad
\beta_X \equiv \sqrt{\frac{|\beta_2|^2 + |\beta_3|^2}{2}},\quad
\beta_\phi \equiv \tau_3-\tau_2.
\label{eq:betas}
\ee
Note that $\beta_3$, $\beta_2$, and $\beta_\phi$ are in general 
complex but $\beta_1$ and $\beta_X$ are real by definition.

In addition to the fluctuations of the gauge fields (\ref{eq:fluctuation_A}), 
we now introduce small fluctuations of the scalar field $T$. 
Since $T$ is a 5 by 5 complex matrix, there are 50 real fluctuations. 
We can separate them in three parts ($\Gamma,\Psi,\Phi$) as
\be
T &=& e^{i \Phi } \left( \tilde T + \Gamma 
+ \left[\Psi,\tilde T\right] \right) e^{- i \Phi } \nonumber\\
&=& \tilde T + \Gamma + \left[\Psi+i\Phi,\tilde T\right] +  \cdots,
\label{eq:T_fluc}
\ee
with
\be
\Gamma =
\begin{pmatrix}
\gamma_3 & 0\\
0 & \gamma_2
\end{pmatrix},\quad
\Psi = \frac{1}{\sqrt2}
\begin{pmatrix}
0 & \psi\\
\psi^\dagger & 0\\
\end{pmatrix}
,\quad
\Phi = 
\frac{1}{\sqrt2}
\begin{pmatrix}
0 & \varphi\\
\varphi^\dagger & 0\\
\end{pmatrix}.
\ee
Here $\gamma_3$ is a 3 by 3 complex matrix and $\gamma_2$ is a 
2 by 2 complex matrix, whereas $\psi$ and $\varphi$ are 3 by 2 
rectangular complex matrices. 
Real degrees of freedom included in $\gamma_3$, $\gamma_2$, $\psi$, 
and $\varphi$ are 18, 8, 12, and 12, respectively.
Summing all them up, we have the correct real degrees of freedom, 
namely 50.
By construction $e^{i\Phi}$ can be regarded as a gauge transformation 
with the broken generator by an amount $\varphi$. 
Hence the fluctuation field $\varphi$ contains the Nambu-Goldstone modes 
corresponding to the broken generators. 
In the following, we keep $\varphi$ and $\psi$ only, and ignore 
$\gamma_{2,3}$ since they are decoupled from the gauge sector or 
they have masses of the order of the GUT scale 
and they do not appear in the low energy effective Lagrangian of the 
gauge fields at the quadratic order.

Let us next substitute ${\cal A}_M$ given in Eq.~(\ref{eq:fluctuation_A}) 
into the first term in the square bracket in Eq.~(\ref{eq:lag}),
and pick up only terms of the quadratic order. 
Those for the unbroken gauge fields can be expressed as 
\be
\lcal_{\alpha} 
&=&
\Tr \bigg[ 
\acal_{\mu}^\alpha \big\{ |\beta_{\alpha}|^{2} \left ( \eta^{\mu \nu} 
\Box - \partial^{\mu} \partial^{\nu} \right ) 
- \eta^{\mu \nu} \partial_{a} |\beta_{\alpha}|^{2} \partial_{a} \big\} 
\acal_{\nu}^\alpha  \nonumber\\
&-& 2 ( \partial^{\mu} \acal_{\mu}^\alpha ) \partial_{a} 
( |\beta_{\alpha}|^{2} \acal_{a}^\alpha ) \nonumber\\
&-& \acal_{a}^\alpha \left ( |\beta_{\alpha}|^{2} \delta_{ab} 
\Box - \delta_{ab} \partial_{c} |\beta_{\alpha}|^{2} \partial_{c} 
+ \partial_{b} |\beta_{\alpha}|^{2} \partial_{a} \right ) \acal_{b}^\alpha 
\bigg],
\label{eq:quadratic_A}
\ee
which is valid regardless of details of the gauge kinetic 
function ${\cal B}$, provided \eqref{eq_T_bg} holds. 
Here $\mu,\nu$ are the four-dimensional spacetime indices, 
$a,b,c$ are used for the extra-dimensional space indices, and 
$\alpha=1,2,3$ are introduced to distinguish $SU(3)$, $SU(2)$, and 
$U(1)$ as $\acal_M^{\alpha=3} = \gcal_M$, $\acal_M^{\alpha=2} = \wcal_M$, 
and $\acal_M^{\alpha=1} = \ycal_M$, respectively.
Note that the trace in Eq.~(\ref{eq:quadratic_A}) for 
${\cal A}_M^{\alpha=1} = \ycal_M$ should be understood as replacing 
$\Tr$ by $\frac{1}{2}$.

The quadratic Lagrangian for the broken gauge field $\xcal_M$ can 
be also obtained in the same way
\be
\lcal_{X} &=&  \Tr \bigg[
 \xcal_{\mu}^{\dagger} \big\{ \beta_X^2 \left ( \eta^{\mu \nu} 
\Box - \partial^{\mu} \partial^{\nu} \right ) 
- \eta^{\mu \nu} \partial_{a} \beta_X^2
 \partial_{a} 
 \big\} \xcal_{\nu} \nonumber \\
&-& 2 ( \partial^{\mu} \xcal_{\mu}^{\dagger} ) 
\partial_{a} \left ( \beta_X^2 \xcal_{a} \right ) \nonumber \\
&-& \xcal_{a}^{\dagger} \left ( \beta_X^2 \delta_{ab} \Box 
- \delta_{ab} \partial_{c} \beta_X^2 \partial_{c} 
+ \partial_{b} \beta_X^2 \partial_{a}
\right ) \xcal_{b}
\bigg].
\label{eq:quadratic_A4}
\ee

Lastly we write down the scalar Lagrangian which can be obtained 
by substituting $T$ given in Eq.~(\ref{eq:T_fluc}) into
$K_\phi = \Tr\left[ D_MT (D^MT)^\dagger\right]$ in Eq.~(\ref{eq:lag}) as
\be
K_\phi
&=& \Tr \bigg\{
\left[ \acal_M+\p_M\Phi , \tilde T\right] 
\left[\acal^M+\p^M\Phi,\tilde T\right]^\dag 
\nonumber\\
&-& \left[\Psi,\tilde T\right]\left(\square - \p_a^2\right) 
\left[\Psi,\tilde T\right]^\dag
+ \left(2i\left[\acal_M+\p_M\Phi,\tilde T\right] 
\left[\p^M\tilde T,\Psi\right] + {\rm h.c.}\right)
\bigg\} \nonumber\\
&=& \Tr \bigg[ 
- \varphi^{\dagger} \left ( |\beta_\phi|^2 \Box 
- \partial_{a} |\beta_\phi|^2 \partial_{a} \right ) \varphi 
+ \varphi^{\dagger} \partial_{a} \left ( |\beta_\phi|^2 \xcal_{a} \right ) 
- |\beta_\phi|^2 \xcal_{a}^{\dagger} \partial_{a} \varphi \nonumber\\
&-& 2 ( \partial^{\mu} \xcal_{\mu}^{\dagger} )  |\beta_\phi|^2 \varphi
+ \xcal_{\mu}^\dagger \eta^{\mu \nu}|\beta_\phi|^2 \xcal_{\nu} 
- \xcal_a^\dagger \delta_{ab}|\beta_\phi|^2\xcal_b\nonumber\\
&-& \psi^{\dagger} \left ( |\beta_\phi|^2 \Box 
- \beta_\phi\partial_{a}^2 \beta_\phi^* \right ) \psi 
+ \left(J_a(\xcal_a^\dag + (\p_a\phi^\dag))\psi + {\rm h.c.}\right)
\bigg],
\ee
where we defined 
\be
J_a = i\left\{\beta_\phi^*(\p_a\beta_\phi) 
- \beta_\phi(\p_a\beta_\phi^*)\right\}.
\label{eq:J}
\ee
We retained $\xcal_\mu$, $\xcal_a$, $\varphi$ and $\psi$, since they 
are mixed with gauge fields. 
However, we omitted $\Gamma$, since it decouples at the quadratic order.
We also need to denote one more quadratic term from the potential 
\be
V_{\phi} = U \Tr \left[ \psi^\dag\psi\right],
\ee
where $U$ is a certain function of the background solution $\tilde T$. 
Here, we also omitted $\Gamma$, since it decouples with $\psi$.
Hence the quadratic Lagrangian is 
\be
\lcal_\phi = K_\phi - V_\phi.
\label{eq:quadratic_A5}
\ee

Finally we introduce a gauge-fixing Lagrangian. 
For the unbroken generators, we introduce the following gauge 
fixing Lagrangian
\be
\lcal^{\rm (gf)}_\alpha = - \dfrac{|\beta_{\alpha}|^{2}}{\xi} 
\Tr \left[ \left(\partial^{\mu} \acal^\alpha_{\mu} 
- \dfrac{\xi}{|\beta_{\alpha}|^{2}} \partial_{a} 
\left( \beta_{\alpha}^{2} \acal^\alpha_{a} \right) \right)^2\right].
\ee
As before $\Tr$ is understood to be replaced by $\frac{1}{2}$ for $\alpha=1$. 
For the broken generators, we introduce another gauge fixing term 
\be
\lcal^{\rm (gf)}_{X\phi} = & - \dfrac{\beta_X^2}{\xi} \Tr \left [
\left(
\partial^{\mu} \xcal_{\mu} - \dfrac{\xi}{\beta_X^2} \left [ \partial_{a} 
\left ( \beta_X^2 \xcal_{a} \right ) + 2 |\beta_\phi|^2 \varphi \right ] 
\right)
\left(
{\rm h.c.}
\right)
\right]
\label{eq:L45}
\ee
Here $\xi$ is an arbitrary gauge fixing constant 
similarly to the $R_\xi$ gauge fixing condition.

\subsection{Compact formulae for the unbroken gauge fields}

\subsubsection{Canonically normalized gauge fields}

The above quadratic Lagrangian is  
complicated and far 
from the standard expression due to the extra $\beta^2$ factor. 
In order to bring it to a more familiar\footnote{
The analysis in this section is a generalization of that in 
Refs.~ \cite{Arai:2012cx,Arai:2013mwa,Arai:2017lfv,Arai:2017ntb,Arai:2018uoy}, 
for the fat brane-world scenario with the domain wall in five dimensions, 
and is a refinement of that in Refs.~\cite{Arai:2018rwf,Eto:2019weg} which also studied 
similar problems in higher dimensions.}
form, let us define 
\be
A_M^\alpha &\equiv& |\beta_\alpha| \acal_M^\alpha,\quad (\alpha = 1,2,3).
\label{eq:cano}
\ee
Below we will also use the expression
$A_M^{\alpha=3} = G_M$, 
$A_M^{\alpha=2} = W_M$, and
$A_M^{\alpha=1} = Y_M$.
In the following we need to deal with the extra-dimensional 
components of gauge field $A_a^\alpha$ differently from 
the four-dimensional fields $A_\mu^\alpha$ due to the fact that 
$\beta_\alpha(x^a)$ depends not on $x^\mu$ but on $x^a$. 
The following vector notation turns out to be 
convenient for describing the low energy effective Lagrangian:
\be
\vec A^{\alpha} \equiv
\begin{pmatrix}
A_4^\alpha\\
A_5^\alpha
\end{pmatrix},\quad \left(\alpha=1,2,3\right) .
\ee

\subsubsection{Vector-analysis-like method}

We now introduce differential operators useful to perform a 
vector-analysis-like method for analyzing mass spectra of gauge fields 
\be
\vec D^{\alpha} \equiv
\begin{pmatrix}
D_4^\alpha\\
D_5^\alpha
\end{pmatrix},\quad \left(\alpha=1,2,3\right)
\label{eq:D_unbroken}
\ee
\be
D^{\alpha}_{a} \equiv - |\beta_\alpha| \partial_{a} \dfrac{1}{|\beta_\alpha|}
= - \p_a + \left(|\beta_\alpha|^{-1}\p_a|\beta_\alpha|\right),
\quad \left(\alpha=1,2,3\right),
\label{eq:def_D}
\ee
where no sum is taken for the index $\alpha$ in the middle and the right-most 
equations.
An adjoint operator of the above differential operator is defined by
\be
\vec D^{\alpha \dagger} = \left(
D_4^{\alpha\dagger},\ D_5^{\alpha\dagger}
\right),
\ee
\be
D_a^{\alpha\dagger} = |\beta_\alpha|^{-1}\p_a|\beta_\alpha| 
= \p_a + \left(|\beta_\alpha|^{-1}\p_a|\beta_\alpha|\right),\quad \left(\alpha=1,2,3\right),
\label{eq:def_Ddag}
\ee
where we do not sum in $\alpha$.

To develop an analogue of the usual vector analysis in three 
spatial dimensions, we introduce analogues of gradient, 
divergence, and rotation in the following way. 
In order to avoid inessential complications, we will suppress 
the index $\alpha$ in the following.
\begin{enumerate}[1)]
\item gradient 
\be
{\rm grad}\,f(x^a) \equiv 
\vec D \circ f(x^a) = 
\begin{pmatrix}
D_4 f(x^a)\\
D_5 f(x^a)
\end{pmatrix}.
\ee 
\item divergence 
\be
{\rm div}\,\vec f(x^a) \equiv 
\vec D^{\dagger} \cdot \vec f(x^a) 
= D_4^{\dagger} f_4(x^a) + D_5^{\dagger} f_5(x^a).
\ee
\item vector rotation 
\be
{\rm rot_v} \vec f(x^a) \equiv \vec D \times \vec f(x^a) 
= D_5 f_4(x^a) - D_4 f_5(x^a).
\ee
\item scalar rotation 
\be
{\rm rot_s} f(x^a) \equiv \vec D^{\dagger} \otimes f(x^a) 
= 
\begin{pmatrix}
D_5^{\dagger} f(x^a)\\ 
- D_4^{\dagger} f(x^a)
\end{pmatrix}.
\ee
\item Laplacian 
\be
\triangle f \equiv {\rm div}\,{\rm grad}\, f 
= \vec D^{\dagger} \cdot \vec D \circ f 
= \sum_{a=4,5} D_a^\dagger D_a f.
\label{eq:Laps}
\ee
\item dual scalar Laplacian 
\be
\tilde\triangle_{\rm s} f \equiv
{\rm rot_v}\,{\rm rot_s}\, f 
= \vec D \times \vec D^{\dagger} \otimes f 
= \sum_{a=4,5}D_a D_a^{\dagger} f.
\label{eq:dLaps}
\ee
\item dual vector Laplacian
\be
\tilde \triangle_{\rm v} \vec f \equiv
{\rm rot_s}\,{\rm rot_v}\, \vec f = \vec{D}^{\dagger} 
\otimes \vec{D} \times \vec f = 
\begin{pmatrix}
D_5^{\dagger}D_5 & - D_5 D_4\\
-D_4^{\dagger} D_5^{\dagger} & D_4^{\dagger}D_4
\end{pmatrix}
\vec f.
\ee
\end{enumerate}
Since $\left[D_4, D_5\right] = 0$ implies 
${\rm rot_v}\,{\rm grad}\, f 
= \vec D \times \vec D \circ f = D_5D_4 f- D_4D_5 f = 0$, 
we find 
\be
{\rm rot_v}\,{\rm grad}\, f = 0.
\label{eq:rot_grad}
\ee
Similarly 
${\rm div}\,{\rm rot_s}\, f =
\vec D^{\dagger}\cdot \vec D^{\dagger} \otimes f =
D_4^{\dagger}D_5^{\dagger} f 
- D_5^{\dagger}D_4^{\dagger} f = 0$ gives 
\be
{\rm div}\,{\rm rot_s}\, f = 0.
\label{eq:div_rot}
\ee
We can define the adjoint of ${\rm div}$ acting on a vector 
function $\vec f(x^a)$ in terms of the inner product between 
the scalar function $h(x^a)$ as 
\be
\int dx^4dx^5\, h^*(x^a) \, {\rm div} \vec f(x^a)
&=& \int dx^4dx^5\, \left( h^* D_4 f_4 + h^* D_5 f_5 \right)\nonumber\\
&=& \int dx^4dx^5\, \left( D_4^\dagger h^*\right)f_4 
+ \left(D_5^\dag h^*\right) f_5 \nonumber\\
&=& \int dx^4dx^5\, \left({\rm grad}\, h(x^a)\right)^\dag \cdot \vec f(x^a),
\ee
where we assumed $\beta(x^a) h(x^a)$ goes to zero at infinity 
$x^a\to\infty$. 
Thus we observe the adjoint of ${\rm div}$ is given by ${\rm grad}$ 
and vice versa 
\be
{\rm grad}^\dagger = {\rm div}
\quad{\rm or}\quad
\left(
\vec D \circ
\right)^\dagger
= \vec D^{\dagger} \cdot\, ,
\ee
\be
{\rm div}^\dagger = {\rm grad}
\quad{\rm or}\quad
\left(
\vec D^{\dagger} \cdot
\right)^\dagger
= \vec D \circ\,.
\label{eq:grad_div}
\ee
Similarly, the adjoint of ${\rm rot_s}$ acting on a scalar function 
$f(x^a)$ can be defined in terms of an inner product with a 
vector function $\vec g(x^a)$ as 
\be
\int dx^4dx^5\, \vec g^\dag \cdot \, {\rm rot_s}  f
&=& \int dx^4dx^5\, \left(g_4 D_5 f - g_5 D_4 f\right) \nonumber\\
&=& \int dx^4dx^5\, \left(D_5^\dag g_4 - D_4^\dag g_5 \right)f \nonumber\\
&=& \int dx^4dx^5\, \left({\rm rot_v}\,\vec g\right)^\dag f.
\ee
Thus we find 
\be
{\rm rot_v}^\dagger = {\rm rot_s}
\quad{\rm or}\quad
\left(
\vec D \times
\right)^\dagger
= \vec D^{\dagger}\otimes\, ,
\ee
\be
{\rm rot_s}^\dagger = {\rm rot_v}
\quad{\rm or}\quad
\left(
\vec D^{\dagger}\otimes
\right)^\dagger
= \vec D \times\,.
\ee
We find also that ${\rm rot_v}\,{\rm grad}$ and 
${\rm div}\,{\rm rot_s}$ are adjoint of each other 
\be
\left({\rm rot_v}\,{\rm grad}\right)^\dagger = {\rm div}\,{\rm rot_s}
\quad{\rm or}\quad
\left(
\vec D \times \vec D \circ
\right)^\dagger
= \vec D^{\dagger}\cdot \vec D^{\dagger} \otimes.
\ee

The final piece of our vector-analysis-like method is a decomposition 
formula for a (two-extra-dimensional) vector field.
Let $\vec A$ be an arbitrary two component vector field. 
There exist scalar fields $B$ and $C$ with which
$\vec A$ is decomposed as
\be
\vec A 
= \,{\rm grad}\, B + \,{\rm rot_s}\, C.
\label{eq:Helmholtz}
\ee 
With the identities in Eqs.~(\ref{eq:rot_grad}) and (\ref{eq:div_rot}), 
this theorem implies that a vector field can be decomposed into 
a rotation-free part and a divergence-free part. 
The theorem is proved in the Appendix~\ref{sec:app1}, and it is 
an analogue of the Helmholtz's theorem for a three-dimensional 
vector field\footnote{
The theorem states that a three-vector field $\vec A$ can be 
decomposed into rotation-free and divergence-free components as
\be
\vec A = \vec\nabla B + \vec\nabla\times\vec C. \nonumber
\ee
}. 
Taking divergence and rotation leads to Poisson-like equations 
\be
\triangle B = {\rm div}\,\vec A, \quad
\tilde \triangle_{\rm s} C = {\rm rot_v}\, \vec A.
\ee
By solving these equations under appropriate boundary conditions, 
we can determine $B$ and $C$ for a given $\vec A$.
Note, however, $B$ and $C$ are not uniquely determined. 
We can determine $B$ and $C$ only up to solutions of Laplace-like 
equations 
\be
\triangle B_0 = 0,\quad 
\tilde \triangle_{\rm s} C_0 = 0.
\ee
Solutions to the above Laplace-like equations are given by 
the kernels of ${\rm grad}\, = \vec D \circ$ and 
${\rm rot_s}\,=\vec D^\dagger \otimes$, respectively. 
From Eqs.~(\ref{eq:def_D}) and (\ref{eq:def_Ddag}), we find 
explicitly that the kernels $B_0, C_0$ are given by 
\be
{\rm grad}\, B_0 = 0 \quad \Rightarrow \quad B_0 \propto |\beta|,\\
{\rm rot_s}\, C_0 = 0 \quad \Rightarrow \quad C_0 \propto |\beta|^{-1}.
\ee
(Remember we have suppressed the index $\alpha = 1,2,3$.)
Hence, when $B_0$ is normalizable, $C_0$ is non normalizable, 
and vice versa. 
On the other hand, 
our Laplacian and the dual Laplacian are positive semi definite, 
as given in Eqs.~(\ref{eq:Laps}) and (\ref{eq:dLaps}).

\subsubsection{Quadratic Lagrangian for unbroken gauge fields}

We can now rewrite the effective Lagrangians 
(\ref{eq:quadratic_A}) in terms of the differential operators 
of the vector-analysis-like method 
\be
\lcal = \Tr \bigg[ A_{\mu} \left( 
\eta^{\mu \nu} \Box - \partial^{\mu} \partial^{\nu} 
+ \eta^{\mu \nu} \triangle \right) A_{\nu}
- 2 \left ( \partial^{\mu} A_{\mu} \right ) 
\left ( {\rm div}\, \vec{A} \right )
- \vec{A}^{\dagger} \left( \Box + \tilde\triangle_{\rm v}\right) 
\vec{A} \bigg],
\label{lag_A}
\ee
after performing a partial integration. 
Next we rewrite the gauge fixing Lagrangians as
\be
{\cal L}^{\rm (gf)} = - \frac{1}{\xi}\Tr\left[\left(\p^\mu A_\mu 
- \xi {\rm div}\,\vec A \right)^2 \right].
\label{eq:A_gf}
\ee
Summing these two up, we get
\be
\lcal + \lcal^{\rm (gf)} &=& 
\Tr \bigg[ A_{\mu} \left( 
\eta^{\mu \nu} \Box - \left(1 - \frac{1}{\xi}\right)
\partial^{\mu} \partial^{\nu} 
+ \eta^{\mu \nu} \triangle \right) A_{\nu} \nonumber \\
&-&  \vec{A}^{\dagger} \left( \Box + \tilde\triangle_{\rm v} 
+ \xi\,{\rm grad}\,{\rm div}\right) \vec{A} \bigg],
\label{eq:Lag_321}
\ee
where we used Eq.~(\ref{eq:grad_div}) 
to obtain 
\be
\left({\rm div} \vec A\right)^\dagger {\rm div} \vec A 
= \vec A^{\dagger}\,{\rm grad}\,{\rm div} \vec A .
\ee

We decompose $\vec A = {\rm grad}\, B + {\rm rot_s}\, C$ using 
the generalized Helmholtz's theorem (\ref{eq:Helmholtz}). 
Note that we can freely exclude the zero mode $B_0 \propto |\beta|$ 
$({\rm grad}\,B_0 = 0$) from $B$ and the zero mode 
$C_0 \propto |\beta|^{-1}$ (${\rm rot_s}\, C_0 = 0$) from $C$ 
without loss of any informations contained in $\vec A$, 
because they never change $\vec A$.
Plugging the decomposition into Eq.~(\ref{eq:Lag_321}), we arrive at 
the following simple formula 
\be
\lcal + \lcal^{\rm (gf)} &=& 
\Tr \bigg[ A_{\mu} \left( 
\eta^{\mu \nu} \Box - \left(1 - \frac{1}{\xi}\right)
\partial^{\mu} \partial^{\nu} 
+ \eta^{\mu \nu} \triangle \right) A_{\nu} \nonumber\\
&-&  
C^{\dagger}\tilde\triangle_{\rm s}
\left(\Box + \tilde\triangle_{\rm s}\right) C
- B^{\dagger} \triangle
\left(\Box + \xi \triangle \right) B
\bigg].
\ee
Since $\triangle$ and 
$\tilde \triangle_{\rm s}$ are semi-positive definite and 
zero modes are excluded from $B$ and $C$, 
we can redefine the divergence-free and rotation-free parts by 
\be
b \equiv \sqrt{\triangle}\, B,\quad
c \equiv \sqrt{\tilde \triangle_{\rm s}}\, C. 
\label{eq:BCbc}
\ee
Substituting these into the last expression, we obtain the 
final formula with the omitted index $\alpha=1,2,3$ retained 
\be
\big(\lcal + \lcal^{\rm (gf)}\big)_{\alpha=1,2,3 } &=& 
\Tr \bigg[ A^\alpha_{\mu} \left( 
\eta^{\mu \nu} \Box - \left(1 - \frac{1}{\xi}\right)
\partial^{\mu} \partial^{\nu} 
+ \eta^{\mu \nu} \triangle^\alpha \right) A^\alpha_{\nu} \nonumber\\
&-&  
c^{\alpha\dagger}
\left(\Box + \tilde\triangle^\alpha_{\rm s}\right) c^\alpha
- b^{\alpha \dagger}
\left(\Box + \xi \triangle^\alpha \right) b^\alpha
\bigg].
\label{eq:Lag_unbroken}
\ee
Note that $b^\alpha$ and $c^\alpha$ are Hermitian matrix 
fields of the same size as $A_\mu^\alpha$, 
since they arise as rotation-free and divergence-free parts 
of $A_{a=4,5}^\alpha$.
We do not take sum over $\alpha$ on the right hand side.

In order to make the Kaluza-Klein (KK) 
expansion more explicit, let us define 
eigenvalues $m_n$ ($\tilde m_n$) and eigenfunctions $B_n(x^a)$ 
($C_n(x^a)$) of $\triangle$ ($\tilde\triangle_{\rm s}$) 
as\footnote{Again, we omit the subscript $\alpha=1,2,3$ from now on.}
\be
\triangle B_n = m_n^2 B_n,
\quad \tilde \triangle_{\rm s} C_n = \tilde m_n^2 C_n.
\label{eq:mode_of_triangle}
\ee
As mentioned before, $\triangle$ and $\tilde \triangle$ are 
positive semi definite. We denote the possible zero eigenvalue 
as $m_{n=0} = \tilde m_{n=0} = 0$, and positive eigenvalues as $n>0$. 
We set the normalizations of the mode functions as usual: 
\be
\int dx^4dx^5\, B_mB_n = \delta_{mn},\quad
\int dx^4dx^5\, C_mC_n = \delta_{mn}.
\ee
Of course, these are meaningful only for normalizable modes \footnote{
More precisely, mode functions should be bounded, so that they can be 
delta-function normalizable for continuum spectra. 
}. 
If $|\beta|$ ($|\beta|^{-1}$) is square integrable, $B_0$ ($C_0$) 
is normalizable but $C_0$ ($B_0$) is not normalizable. 

Now, we decompose $B$ ($C$) as
\be
B(x^M) = \sum_{n} f_n(x^\mu) B_n(x^a),\quad
C(x^M) = \sum_{n} g_n(x^\mu) C_n(x^a),
\ee
where expansion coefficients $f_n$ and $g_n$ are effective fields 
in four dimensions, and the sum is taken only for the 
normalizable modes. 
Eq.~(\ref{eq:BCbc}) implies that expansions for $b$ and $c$ do 
not have zero modes 
\be
b(x^M) = \sum_{n>0} b^{(n)}(x^\mu) B_n(x^a),\quad
c(x^M) = \sum_{n>0} c^{(n)}(x^\mu) C_n(x^a),
\label{eq:expand_bc}
\ee
with $b^{(n)} = f_n m_n$ and $c^{(n)} = g_n \tilde m_n$, since 
zero modes are excluded in defining $B$ and $C$ from $\vec A$. 

In contrast, the four-dimensional components do have the zero 
mode of $\triangle$ because there is no reason to eliminate it
\be
A_\mu(x^M) = \sum_{n \ge 0} A_\mu^{(n)}(x^\mu) B_n(x^a).
\label{eq:expand_A}
\ee
Since we wish to have the massless gauge fields localized on 
the vortices, we should impose the square integrable condition on $\beta$
(This is one of the most important results of this work) as
\be
\frac{1}{g^2} \equiv \int dx^4dx^5\, |\beta|^2 < \infty.
\label{eq:gauge-coupling}
\ee
This gives the normalization of the mode function for $n=0$ as
\be
B_0(x^a) = g |\beta(x^a)|.
\ee
Substituting the expansions in Eqs.~(\ref{eq:expand_bc}) and 
(\ref{eq:expand_A}) into Eq.~(\ref{eq:Lag_unbroken}), 
and integrating it over the $x^4$-$x^5$ plane, we obtain the 
effective Lagrangians for the KK towers.

The massless mode ($n=0$) only appears in the sector of the 
four-dimensional gauge fields,
\be
{\cal L}_\alpha^{(0)} = 
\Tr \left[ A^{\alpha(0)}_{\mu} \left( 
\eta^{\mu \nu} \Box - \left(1 - \frac{1}{\xi}\right)
\partial^{\mu} \partial^{\nu} \right) A^{\alpha(0)}_{\nu}\right].
\ee
This is nothing but the quadratic Lagrangian of massless vector 
fields. In order to recover self-interaction terms of 
non-Abelian gauge fields, 
we return to Eq.~(\ref{eq:cano}). We have 
$A_\mu^\alpha(x^M) = B_0^\alpha(x^a) A_\mu^{\alpha(0)}(x^\mu) + \cdots$, 
therefore
\be
{\cal A}_\mu^\alpha(x^M) = \beta_\alpha^{-1}(x^a) 
A_\mu^\alpha(x^M) = g_\alpha A_\mu^{\alpha(0)}(x^\mu) + \cdots.
\label{eq:calA_exp}
\ee
It is remarkable that the mode function of the zero mode with 
respect to the original field ${\cal A}_\mu$ is constant namely $g_\alpha$. 
This ensures the universality of the gauge coupling constant. 
Nevertheless, the zero mode effective Lagrangian is well-defined 
because of the extra factor ${\cal B}(T)\bcal(T)^\dag$ in 
Eq.~(\ref{eq:lag}) gives the necessary suppression factor 
$|\beta(x^a)|^2$ in the extra dimensions. 
Keeping only zero mode in Eq.~(\ref{eq:calA_exp}), the field 
strength reads 
\be
{\cal F}_{\mu\nu}^\alpha &=& g_\alpha F_{\mu\nu}^{\alpha(0)}, \\
\quad
F_{\mu\nu}^{\alpha(0)} &=& \p_\mu A_\nu^{\alpha(0)} 
- \p_\nu A_\mu^{\alpha(0)} + i g_\alpha \left[A_\mu^{\alpha(0)},
A_\nu^{\alpha(0)}\right].
\label{eq:F_4d}
\ee
This should be compared with the original field strength 
${\cal F}_{MN} = \p_M {\cal A}_N - \p_N {\cal A}_M 
+ i \left[{\cal A}_M,{\cal A}_N\right]$ in which gauge coupling constant
is absorbed in the gauge field ${\cal A}_M$.
The four-dimensional effective gauge coupling appears in 
Eq.~(\ref{eq:F_4d}) because of the normalization condition in 
\eqref{eq:gauge-coupling}. 
One can also understand this result as a consequence of unbroken 
four-dimensional local gauge invariance\cite{Ohta:2010fu}. 
Now, including the self interactions of the non-Abelian gauge fields, 
the zero mode effective Lagrangian is found to be the standard one as
\be
\int dx^4dx^5\, \Tr\left[-\frac{\beta_\alpha^2}{2}g_\alpha^2 
F_{\mu\nu}^{\alpha(0)}F^{\alpha(0)\mu\nu} \right]
= - \frac{1}{2} \Tr\left[F_{\mu\nu}^{\alpha(0)}F^{\alpha(0)\mu\nu} \right].
\ee

We would like to emphasize a new important feature of our model compared 
to many previous works. 
The gauge kinetic function $\bcal\bcal^\dag$ in Eq.~(\ref{eq:lag}) 
is not a scalar but a matrix as a nontrivial representation of 
gauge group. 
A similar mechanism of localizing massless gauge fields on 
topological solitons have been studied, which utilize a conformal factor.
However, it is usually a singlet of the gauge group, since it 
usually originates from a spacetime metric or dilaton. 
Such a singlet conformal factor cannot distinguish various components 
of $SU(5)$ gauge fields, unlike our model.

As for the higher KK modes with $n>0$, we have two separated parts. 
The one is for massive vector fields $A^{\alpha(n)}_{\mu}$
\be
{\cal L}_{1,\alpha}^{(n>0)} &=& 
\Tr \bigg[ A^{\alpha(n)}_{\mu}
\left( 
\eta^{\mu \nu} \Box - \left(1 - \frac{1}{\xi}\right)\partial^{\mu} 
\partial^{\nu} 
+ \eta^{\mu \nu} (m_n^{\alpha})^2 \right) A^{\alpha(n)}_{\nu} \nonumber\\
&-& b^{\alpha(n) \dagger}
\left(\Box + \xi (m_n^{\alpha})^2 \right) b^{\alpha(n)}
\bigg].
\ee
Note that the vector field $A_\mu^{\alpha(n)}$ has two components, 
with the mass squared $(m_n^\alpha)^2$ and $\xi (m_n^\alpha)^2$. 
On the other hand, the scalar $b$ is originated from the rotation-free 
part of $\vec A^\alpha$, and has the mass squared $\xi (m_n^\alpha)^2$. 
It should be combined with the above components of vector field 
$A_\mu^{\alpha(n)}$ with the same mass squared together with 
the ghost and anti-ghost fields to become unphysical, as can be 
recognized from their gauge dependent mass. 
This situation is analogous to the usual situation in the $R_\xi$ gauge. 
Thus we see that the rotation free part plays a role of an (unphysical) 
Nambu-Goldstone field to give a mass to the KK tower of vector fields. 
The other part includes the divergence free part $c$:
\be
{\cal L}_{2,\alpha}^{(n>0)} = -
\Tr \left[ c^{\alpha(n)\dagger}
\left(\Box + (\tilde m_n^\alpha)^2 \right) c^{\alpha(n)}
\right].
\ee
This should be a physical scalar fields with the mass squared 
$(m_n^\alpha)^2$. 

Summary of the unbroken sector $SU(3)\times SU(2) \times U(1)$ 
is the following:
The six-dimensional gauge fields ${\cal A}_M^\alpha$ provide one 
massless gauge field $A_\mu^{\alpha(0)}$,
KK tower of massive vector fields $A_\mu^{\alpha(n>0)}$, and 
KK tower of massive scalar fields
$c^{\alpha(n>0)}$ to the four-dimensional effective theory.

\subsection{Compact formulae for the broken gauge fields}

We move to the broken sector with $\xcal_M$ and $\varphi$ which is 
more complicated than the unbroken sector in the previous subsection.
We will make the effective Lagrangians (\ref{eq:quadratic_A4}), 
(\ref{eq:quadratic_A5}), and (\ref{eq:L45}) as simple as possible.

Firstly, we define canonically normalized fields by
\be
X_M \equiv \beta_X \xcal_M,\quad
X_6 \equiv |\beta_\phi| \varphi,\quad
X_7 \equiv |\beta_\phi| \psi.
\ee
We will also use the notation
$X_6 = \phi$ later.
We can treat the NG boson $\varphi$ and $\psi$ as if they are 
a sixth and a seventh components of a massive gauge field, by 
defining a four-component vector as 
\be
\bm{X} \equiv
\begin{pmatrix}
X_4\\
X_5\\
X_6\\
X_7
\end{pmatrix},
\ee
which will bring a benefit of simplification in many expressions below. 
Assigning the NG boson $X_6 = \phi$ to be the sixth gauge field is somehow natural 
because would-be eaten NG bosons can be regarded as a part of the massive vector field in general.
As before, let us introduce the differential operators by
\be
\vec D^{X} &\equiv& - \beta_X \vec \partial \dfrac{1}{\beta_X}
= - \vec \p + \left(\beta_X^{-1}\vec \p\beta_X\right),\\
\vec D^{\phi} &\equiv& - |\beta_\phi| \vec\partial \dfrac{1}{|\beta_\phi|}
= - \vec\p + \left(|\beta_\phi|^{-1}\vec\p |\beta_\phi|\right).
\ee
Note that $\vec D^X$ can be expressed by $\vec D^\phi$ by  
\be
\vec D^X \circ = \mcal^{-1} \vec D^\phi \circ \mcal,\quad
\vec D^\phi \circ = \mcal \vec D^X \circ \mcal^{-1},\qquad
\mcal \equiv \frac{|\beta_\phi|}{\beta_X}.
\label{eq:DXP}
\ee
Then we define a four-component (effectively three-component) operator by
\be
\bm{D} \equiv
\begin{pmatrix}
\vec D^X\\
\mcal\\
0
\end{pmatrix}.
\label{eq:Dop_extra}
\ee
From this we can define 
\begin{enumerate}[1)]
\item gradient
\be
{\rm Grad}\,f \equiv 
\bm{D} \circ f = 
\begin{pmatrix}
\vec D^X\\
\mcal\\
0
\end{pmatrix}
\circ
f = 
\begin{pmatrix}
\vec D^X \circ f\\
0\\
0
\end{pmatrix}+
\begin{pmatrix}
\vec 0\\
\mcal f\\
0
\end{pmatrix}.
\ee 
\item divergence
\be
{\rm Div}\,\bm{f} \equiv
\bm{D}^\dag \cdot \bm{f} = 
\left(
\vec D^{X\dag} ,\ 
\mcal ,\
0
\right) \cdot 
\begin{pmatrix}
\vec f\\
f_6 \\
f_7
\end{pmatrix}
= 
\vec D^{X\dag} \cdot \vec f +
\mcal f_6.
\ee 
\item Laplacian
\be 
\bm{\Delta} f \equiv {\rm Div}\,{\rm Grad}\,f = \bm{D}^\dagger\cdot\bm{D}\circ f = \triangle^X f + \mcal^2 f
\label{eq:LAP}
\ee
with $\triangle^X = {\rm div}^X\,{\rm grad}^X\, = \vec D^{X\dag} \cdot \vec D^X \circ$.
Note that $\bm{\Delta}$ is positive definite since $\triangle^X$ is non-negative definite and $\mcal^2$
is positive definite. 
\item dual vector Laplacian
\be
\tilde{\bm{\Delta}}_{\rm v} \equiv
\left(
\begin{array}{ccc}
\tilde{\triangle}_{\rm v}^X\,+\mcal^2{\bf 1}_2 & - \mcal \vec D_\phi\circ & - \frac{1}{|\beta_\phi|^2}\mcal \vec J\\
- \vec D_\phi^\dag \cdot \mcal & \triangle^\phi & \vec D^{\phi\dag} \cdot \vec J \frac{1}{|\beta_\phi|^2}\\
- \frac{1}{|\beta_\phi|^2}\mcal \vec J^\dag & \ \frac{1}{|\beta_\phi|^2}\vec J^\dag \vec D^\phi\circ\  &
\ \beta_\phi \vec D^{\phi\dag}\cdot \frac{1}{|\beta_\phi|^2}\vec D^\phi \circ \beta_\phi^* + U
\end{array}
\right).
\label{eq:tildeDelta_v}
\ee
with the 2 by 2 Laplacian
$\tilde{\triangle}_{\rm v}^X = \vec D^{X\dag} \otimes \vec D^X \times$ 
and the 1 by 1 Laplacian $\triangle^\phi = \vec D^{\phi\dag} \cdot \vec D^\phi \circ$.
The elements of two component vector $\vec J$ are given by Eq.~(\ref{eq:J}).
Note that $\tilde{\bm{\Delta}}_{\rm v}$ itself is 4 by 4.
\end{enumerate}

It would be nice if we could factorize $\tilde{\bm{\Delta}}_{\rm v}$ as 
$\tilde{\bm{\Delta}}_{\rm v} = \,{\rm Rot_s}\,{\rm Rot_v}\,$ 
with appropriate rotation operators ${\rm Rot_s}$ and ${\rm Rot_v}$.
Unfortunately, this does not seem very easy, so we abandon it.
Nevertheless, we still have identities corresponding to
${\rm Rot_v}\,{\rm Grad}\, = \bm{0}$ and ${\rm Div}\,{\rm Rot_s}\, = 0$:
\be
\tilde{\bm{\Delta}}_{\rm v}\,{\rm Grad}\, = \bm{0},
\label{eq:id3}
\ee
and 
\be
{\rm Div}\,\tilde{\bm{\Delta}}_{\rm v} = 0.
\label{eq:id4}
\ee
The former is verified as
\be
\tilde{\bm{\Delta}}_{\rm v}\,{\rm Grad}\,f = 
\tilde{\bm{\Delta}}_{\rm v}\,
\begin{pmatrix}
\vec D^X \circ f\\
\mcal f\\
0
\end{pmatrix}
= \begin{pmatrix}
(\tilde{\triangle}_{\rm v}^X+\mcal^2) \vec D^X \circ f - \mcal \vec D^\phi \circ \mcal f \\
-\vec D^{\phi\dag}\cdot\mcal\vec D^X\circ f + \vec D^{\phi\dag}\cdot\vec D^\phi\circ\mcal f\\
- \frac{1}{|\beta_\phi|^2}\mcal \vec J^\dag \vec D^X\circ f + 
\frac{1}{|\beta_\phi|^2}\vec J^\dag \vec D^\phi\circ\mcal f
\end{pmatrix}
= \bm{0},
\ee
where we used a similar identity 
$\tilde{\triangle}_{\rm v}^X\vec D^X \circ = \vec D^{X\dag} \otimes (\vec D^X \times\vec D^X \circ) = 0$ 
to Eq.~(\ref{eq:rot_grad}) for $\vec D^X$ and Eq.~(\ref{eq:DXP}).
The latter can also be verified as
\be
{\rm Div}\,\tilde{\bm{\Delta}}_{\rm v}\bm{f} &=& 
{\rm Div}\,
\begin{pmatrix}
(\tilde{\triangle}_{\rm v}^X\,+\mcal^2) \vec f - \mcal \vec D_\phi\circ f_6 - \frac{1}{|\beta_\phi|^2}\mcal \vec J f_7\\
- \vec D_\phi^\dag \cdot \mcal \vec f + \triangle^\phi f_6 +  \vec D^{\phi\dag} \cdot \vec J \frac{1}{|\beta_\phi|^2} f_7\\
- \frac{1}{|\beta_\phi|^2}\mcal \vec J^\dag \vec f + \ \frac{1}{|\beta_\phi|^2}\vec J^\dag \vec D^\phi\circ\  f_6
+ (\beta_\phi \vec D^{\phi\dag}\cdot \frac{1}{|\beta_\phi|^2}\vec D^\phi \circ \beta_\phi^* + U)f_7
\end{pmatrix} \nonumber\\
&=& \vec D^{X\dag}\cdot \left((\tilde{\triangle}_{\rm v}^X\,+\mcal^2) \vec f - \mcal \vec D_\phi\circ f_6 - \frac{1}{|\beta_\phi|^2}\mcal \vec J f_7\right) \nonumber\\
&+& 
\mcal\left(- \vec D_\phi^\dag \cdot \mcal \vec f + \triangle^\phi f_6 +  \vec D^{\phi\dag} \cdot \vec J \frac{1}{|\beta_\phi|^2} f_7\right) \nonumber\\
&=& 0,
\ee
where we used a similar identity
$\vec D^{X\dag}\cdot \tilde{\triangle}_{\rm v}^X = (\vec D^{X\dag}\cdot \vec D^{X\dag} \otimes) \vec D^X \times = 0$
to Eq.~(\ref{eq:div_rot}) for $\vec D^X$ and Eq.~(\ref{eq:DXP}).

Since we did not succeed to define appropriate rotation operators 
in the case of broken generators, we cannot introduce a Helmholtz-like 
decomposition for the four-vector $\bm{X}$. 
Instead, we introduce a projection operator 
\be
\bm{P} = \,{\rm Grad}\,\bm{\Delta}^{-1}\, {\rm Div}.
\ee
This satisfies $\bm{P}^2 = \bm{P}$, and is well-defined because $\bm{\Delta}$ is positive definite. 
We decompose $\bm{X}$ as
\be
\bm{X} =  \bm{P} \bm{X} + (1-\bm{P})\bm{X}.
\label{eq:HHD_3}
\ee
The first term is ``rotation-free'' because
\be
\bm{P} \bm{X} = \,{\rm Grad}\, Y,\quad \bm{\Delta} Y = {\rm Div}\,\bm{X}.
\label{eq:Hel_X}
\ee
The second term is divergence-free because
\be
{\rm Div}\,(1-\bm{P})\bm{X} = \,{\rm Div}\,\bm{X} - {\rm Div}\,\left({\rm Grad}\,\bm{\Delta}^{-1}\,{\rm Div}\,\right)\bm{X} = 0.
\ee
The Hermitian conjugate of this is $(1-\bm{P})\,{\rm Grad}\, = 0$.

Now, we are ready to rewrite the Lagrangians  (\ref{eq:quadratic_A4}), 
(\ref{eq:quadratic_A5}), and (\ref{eq:L45}) in more compact forms.
Let us first rewrite (\ref{eq:quadratic_A4}) and (\ref{eq:quadratic_A5})
\be
{\cal L}_X + {\cal L}_\phi &=& \Tr\bigg[
X_\mu^\dagger\left(
\eta^{\mu\nu}\square - \p^\mu\p^\nu + \eta^{\mu\nu} \bm{\Delta}\right) X_\nu \nonumber\\
&-& 2 \left(\p^\mu X_\mu^\dagger\right)\,{\rm Div}\,\bm{X}
- \bm{X}^\dagger \left(\square + \tilde{\bm{\Delta}}_{\rm v}\right) \bm{X}
\bigg].
\label{eq:lag_X}
\ee
Remarkably, unification of $\vec X$ originated from the extra-dimensional component of $\xcal_M$, the NG boson $X_6=\phi$ and the 
additional scalar $X_7$
into the four-component vector $\bm{X}$ is essential to have the quadratic Lagrangian in a compact form.
Furthermore, \eqref{eq:lag_X} for $X_\mu$ and $\bm{X}$ is formally 
identical to Eq.~(\ref{lag_A}) for the unbroken gauge field $A_\mu$ 
and $\vec A$.

Similarly, the gauge fixing Lagrangian (\ref{eq:L45}) can be expressed as
\be
\lcal^{\rm (gf)}_{X\phi} = -\frac{1}{\xi} \Tr\,\left[
\left(\p^\nu X_\nu - \xi \,{\rm Div}\,\bm{X}\right)^\dag
\left(\p^\mu X_\mu - \xi \,{\rm Div}\,\bm{X}\right)
\right].
\label{eq:X_gf}
\ee
This is a counterpart of Eq.~(\ref{eq:A_gf}).

Adding Eqs.~(\ref{eq:lag_X}) and (\ref{eq:X_gf}), 
we have the following quadratic Lagrangian
\be
{\cal L}_X + {\cal L}_\phi + \lcal^{\rm (gf)}_{X\phi} &=&
\Tr\bigg[
X_\mu^\dagger\left(
\eta^{\mu\nu}\square - \left(1-\frac{1}{\xi}\right) \p^\mu\p^\nu + \eta^{\mu\nu} \bm{\Delta}\right) X_\nu \nonumber\\
&-& \bm{X}^\dagger \left(\square + \tilde{\bm{\Delta}}_{\rm v} + \xi\,{\rm Grad}\,{\rm Div}\, \right) \bm{X}
\bigg],
\label{eq:Lag_Xa}
\ee
where we used
\be
\left({\rm Div}\,\bm{X}\right)^\dag \left({\rm Div}\,\bm{X}\right) = 
\bm{X}^\dag\,{\rm Grad}\,{\rm Div}\,\bm{X}.
\ee
This is clearly a counterpart of Eq.~(\ref{eq:Lag_321}).

Thus we accomplished obtaining much simpler formula compared to the initial one.
But the most important point is that Eq.~(\ref{eq:Lag_Xa}) for the broken gauge field with the NG fields
is expressed in the form which is formally identical in fashion found in Eqs.~(\ref{eq:Lag_321}) for the unbroken gauge fields.
Since we have developed the way to extract physical spectra for the latter, we just formally but partially repeat the same procedures.

Therefore, what we have to do next is to decompose $\bm{X}$ as 
$\bm{X} = {\rm Grad}\, Y + (1-\bm{P})\bm{X}$. 
Substituting this into Eq.~(\ref{eq:Lag_Xa}) and using the identities Eqs.~(\ref{eq:id3}) and (\ref{eq:id4}), we get
\be
{\cal L}_X + {\cal L}_\phi + \lcal^{\rm (gf)}_{X\phi} &=&
\Tr\bigg[
X_\mu^\dagger\left(
\eta^{\mu\nu}\square - \left(1-\frac{1}{\xi}\right) \p^\mu\p^\nu + \eta^{\mu\nu} \bm{\Delta}\right) X_\nu \nonumber\\
&-& \left((1-\bm{P})\bm{X}\right)^\dag\left(\square + \tilde{\bm{\Delta}}_{\rm v}\right)(1-\bm{P})\bm{X}
- Y \bm{\Delta} \left(\square + \xi \bm{\Delta} \right) Y
\bigg].
\ee
The last treatment is making this canonical by redefining $Y$ and $(1-\bm{P})\bm{X}$ by
\be
y = \sqrt{\bm{\Delta}}\,Y,\quad
\bm{x} = (1-\bm{P})\bm{X}.
\ee
Substituting this into the above expression, we get the final form of the quadratic Lagrangian
of the broken sector
\be
{\cal L}_X + {\cal L}_\phi + \lcal^{\rm (gf)}_{X\phi} &=&
\Tr\bigg[
X_\mu^\dagger\left(
\eta^{\mu\nu}\square - \left(1-\frac{1}{\xi}\right) \p^\mu\p^\nu + \eta^{\mu\nu} \bm{\Delta}\right) X_\nu \nonumber\\
&-& \bm{x}^\dagger  \left(\square  + \tilde{\bm{\Delta}}_{\rm v}\right) \bm{x}
- y  \left(\square + \xi \bm{\Delta} \right) y
\bigg].
\label{eq:eff_lag_X}
\ee

We have seen remarkable similarities between the unbroken and broken sectors.
From now on, we will shed light on differences between them.
Firstly, the Laplacian $\bm{\Delta}$ given in Eq.~(\ref{eq:LAP}) is positive definite.
(This should be compared with $\triangle^\alpha$ in Eq.~(\ref{eq:Lag_unbroken}) which is only non-negative.)
This leads to an important physical consequence that there are no massless modes in the four-dimensional
component of the broken gauge field $X_\mu$. This is, of course, expected from the beginning because
$X_\mu$ corresponds to the gauge fields associated with the broken generators.
What is not totally trivial here is to clarify which modes are eaten by $X_\mu$.
The answer is $y$. We are lead to this conclusion from the fact that $X_\mu$ and $y$ share the same 
mass operator $\bm{\Delta}$ except for the extra constant factor $\xi$ for $y$. Again, we come across
a generalized form of the $R_\xi$ gauge
(This should be compared to the relation between $\vec A_\mu$ and $b$ in Eq.~(\ref{eq:Lag_unbroken})).
Appearance of $\xi$ implies that $y$ plays a role of infinite tower of the NG modes which are 
absorbed by infinite tower of $X_\mu$ including the bottom of the tower. 
Remembering the definition given in Eq.~(\ref{eq:Hel_X}), the source for $y$ is 
${\rm Div}\,\bm{X} = \vec D^{X\dag} \cdot \vec X + \mcal \phi$. Therefore, the broken gauge field eats 
not pure NG modes $\phi$ but 
a mixture of the rotation-free part $\vec D^{X\dag} \cdot \vec X$ and the NG modes $\phi$.
Specifying the lowest mass of $\bm{\Delta}$ is in general difficult except for a very special case
where $\mcal$ is a constant. If $\mcal$ is a constant, $\bm{\Delta} = \triangle^X + \mcal^2$ 
and $\triangle^X$ differ by just a constant 
$\mcal^2$.  We know that the zero mode of the non-negative definite operator $\triangle^X$ 
is given by $\beta_X$. Therefore, the lowest mass is the constant $\mcal$ itself. 
However, in general $\mcal$ is not constant, so we are only sure that there is a finite mass gap but 
explicit spectrum of $\bm{\Delta}$ depends on $\bcal$.

Our final comment is on the divergence-free component $\bm{x}$.
As we mentioned above, we were not able to factorize the Laplacian 
$\tilde{\bm \Delta}_{\rm v}$ as a product of
rotation operators. Remembering the unbroken part where we decomposed 
$\vec A$ by the Helmholtz-like decomposition,
and we succeeded in splitting the two component vector into the 
two scalar $b$ and $c$ by replacing $\tilde{\triangle}_{\rm v} 
= \,{\rm rot}_{\rm s}\,{\rm rot}_{\rm v}$ by 
$\tilde{\triangle}_{\rm s} = \,{\rm rot}_{\rm v}\,{\rm rot}_{\rm s}$ 
as in Eq.~(\ref{eq:Lag_unbroken}).
Furthermore, this rewriting allows us to conclude that the 
divergence-free component $c$ does not have the zero mode of 
the non-negative operator $\tilde\triangle_{\rm s}$.
At present unfortunately, we cannot make a similar statement 
for $\tilde{\bm{\Delta}}_{\rm v}$. 
We are sure that there are no negative eigenvalues, but 
we cannot exclude a zero eigenvalue. It is an important open 
problem.
But if we could do, it would not so useful in the following sense. 
Recall $\tilde{\bm{\Delta}}_{\rm v}$ acts on the four-component 
vector $(1-\bm{P})\bm{X}$.
Precisely speaking, since the divergence part is projected out, 
the degrees of freedom of the four-component vector $(1-\bm{P})\bm{X}$
is not four but three.
So if we succeeded in dualizing $\tilde{\bm{\Delta}}_{\rm v}$, 
we are still left with the three-component vector.
It is a complicated problem to find spectrum of the 4 by 4 matrix 
operator $\tilde{\bm{\Delta}}_{\rm v}$ given in Eq.~(\ref{eq:tildeDelta_v}), 
since it highly depends on the details of models. 
We leave the spectrum of this operator as one of future problems.

\subsection{A typical example}

Let us illustrate our models more explicitly by a concrete choice 
of $\bcal$ as an example. 
One of the simplest choice of $\bcal$ is
\be
\bcal\bcal^\dag = \left(v^2{\bf 1}_5 - T^\dagger T\right)^2.
\label{eq:emp_B}
\ee
Note that this is just a single mere one possibility. One can consider, say,
$\bcal\bcal^\dag = \left(v^2{\bf 1}_5 - T^\dagger T\right)^s$ with $s\ge2$. 
The crucial is if the condition (\ref{eq:gauge-coupling}) is satisfied or not.
We will verify that for the above simplest choice it is.
$T$ is with the 3-2 splitting vortex background solution 
(see also Eq.~(\ref{eq:3-2}))
\be
T = \left(\begin{array}{cc}\tau_3{\bf 1}_3 & \\ & 
\tau_2{\bf 1}_2\end{array}\right),
\quad 
\tau_2 = v f_2(r_2)e^{i\theta_2},\quad
\tau_3 = v f_3(r_3)e^{i\theta_3}.
\ee
For simplicity, we consider the minimal winding number, namely 
each diagonal component has a single winding number. 
Here $(r_2,\theta_2)$ and $(r_3,\theta_3)$ are two dimensional 
polar coordinates defined as $x^4-a_2+i(x^5-b_2) = r_2 e^{i\theta_2}$ 
and $x^4-a_3+i(x^5-b_3) = r_3 e^{i\theta_3}$ where 
$(a_3,b_3)$ is the position of the vortex associated to $SU(3)$ 
and $(a_2,b_2)$ is that of the vortex associated to $SU(2)$.
We impose the profile functions to satisfy the boundary conditions 
$f_{2,3}(0) = 0$ and $f_{2,3}(\infty) = 1$. 
Solutions of the vortex equation have been obtained only numerically, 
but we do not need precise solutions for $f_2$ and $f_3$ 
in order to understand qualitative aspects of gauge field localization. 
Hence, we use the following approximation 
\be
f_\alpha(r) = \frac{r_\alpha}{\sqrt{1+r_\alpha^2}},\quad (\alpha=2,3).
\ee
Note that these satisfy not only the correct boundary condition 
but also exhibit a good asymptotic behavior $f_{2,3} \to r_{2,3}$ 
($r_{2,3}\to0$) and $f_{2,3} \to 1 - 1/(2r_{2,3}^2)$ 
($r_{2,3} \to \infty$) as a global vortex. 
Then we have
\be
|\beta_\alpha(r)| = v^2\left(1-f_\alpha(r_\alpha)^2\right) 
= \frac{v^2}{1+r_\alpha^2},\quad (\alpha=2,3),
\ee
which satisfy the square integrability condition (\ref{eq:gauge-coupling}) because their 
asymptotic behaviors are $|\beta_{2,3}| \to {v^2}/{r^2}$ as 
$r_{2,3} \sim r \to \infty$.\footnote{
One might naively expect that the unbroken $H = SU(3)\times SU(2) \times U(1)$ gauge fields
remain massless everywhere for the scalar field profile 
of the global vortex only reduces as a power low. However, this is not correct.
Localization of the $H$ gauge fields does not depend on the scalar profile itself but the 
gauge kinetic function $\beta$ ($\bcal$). Even though our background solution is the global
vortex, it can localize the massless non-Abelian gauge fields by the square integrable $\beta$.
Of course, if one considers a local vortex instead of the global one, satisfying the square
integrability condition becomes much easier, so that range of possibilites for $\bcal$ would become larger.
We will investigate the case of a local vortex elsewhere.
}
Now the corresponding $\triangle^{\alpha}$'s for $\alpha=2,3$ 
are given by 
\be
\triangle^{\alpha} &=& \sum_{a=4,5} \left(-\p_a^2 
+ \frac{\p_a^2|\beta_\alpha|}{|\beta_\alpha}|\right)
= - \nabla^2 + \vcal_\alpha(r),\quad \vcal_\alpha(r) 
= \frac{4(r_\alpha^2-1)}{(1+r_\alpha^2)^2},\\
\tilde{\triangle}_{\rm s}^{\alpha} &=& \sum_{a=4,5} 
\left(-\p_a^2 + \frac{\p_a^2|\beta_\alpha|^{-1}}{|\beta_\alpha|^{-1}}\right)
= - \nabla^2 + \tilde{\vcal}_\alpha(r),\quad \tilde{\vcal}_\alpha(r)
= \frac{4}{1+r_\alpha^2}.
\ee
The eigenvalue problems for these operators are mere two-dimensional 
Schr\"odinger equations with the axially symmetric potentials 
$\vcal_\alpha$ and $\tilde\vcal_\alpha$. 
The normalizable zero mode of $\vcal_\alpha$ is $\beta_\alpha(r)$ 
itself and it is the only discrete spectrum\footnote{
Of course, this is merely a property of the specific choice of 
$\bcal^2$ in Eq.~(\ref{eq:emp_B}), and not a generic property for 
other possible choices.} of $\triangle^\alpha$.
On the other hand, there is no discrete spectrum such as normalizable 
zero modes for $\tilde{\triangle}^\alpha$ since $\tilde{\vcal}_\alpha$ 
is a positive convex function.
This means that massless four-dimensional gauge fields $G_\mu$ 
and $W_\mu$ exist only in the $SU(3)$ and $SU(2)$ gauge group. 
For this particular choice of $\bcal$, there exists a massive 
localized modes as well in the four-dimensional vector component 
in the adjoint representation of the $SU(3)$ and $SU(2)$ gauge group. 
However, together with the continuum spectra in both four-dimensional 
and extra-dimensional components, all these massive modes have 
an energy gap of order the GUT scale. 

\begin{figure}[h]
\begin{center}
\includegraphics[width=15cm]{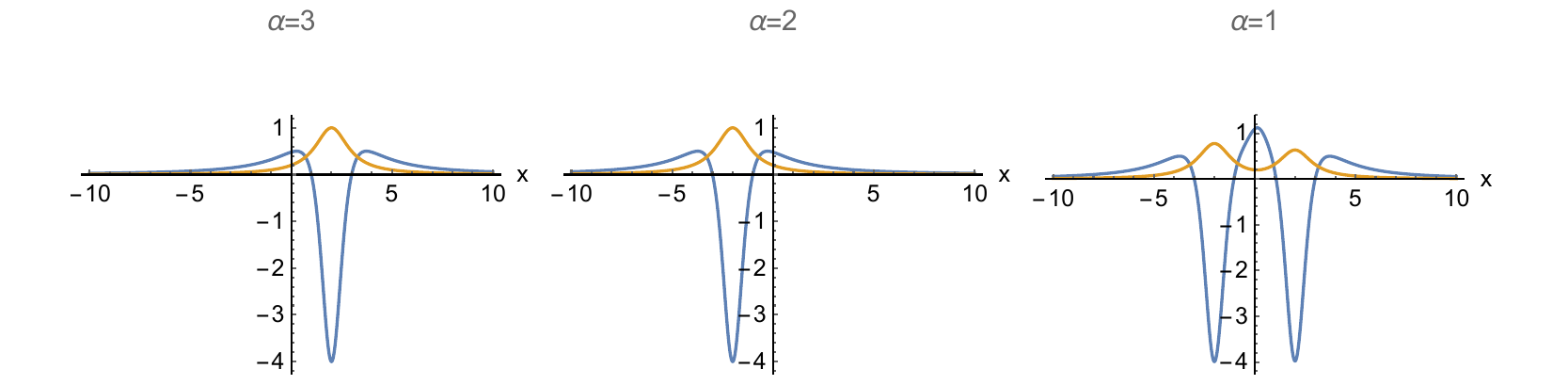}
\caption{$\vcal_\alpha$ and the zero mode $\beta_\alpha$ for 
$\alpha=3,2,1$. We set $(a_3,b_3) = (2,0)$
and $(a_2,b_2) = (-2,0)$. The figures show cross sections on the 
$x^4$ axis ($x^5=0$). The left-most panel shows the
gluon $G_\mu$, the middle for the weak boson $W_\mu$, and the 
right-most for the hyper $U(1)$ $Y_\mu$.}
\label{fig:example_321}
\end{center}
\end{figure}

\begin{figure}[ht]
\begin{center}
\includegraphics[width=14cm]{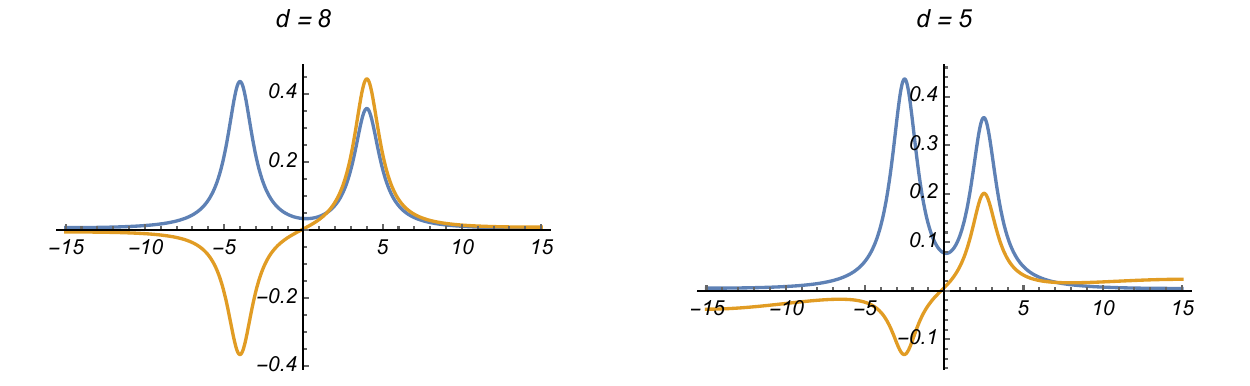}
\caption{
The persistent zero mode $\beta_1$ (blue) and the 
lifted zero mode (orange).
The positions of the 3- and 2-vortices are set to be 
$(a_3,b_3) = (d/2,0)$ and $(a_2,b_2) = (-d/2,0)$, and 
the left figure is with $d=8$ and the right one is with $d=5$. 
The lifted zero mode in the left graph is normalizable while 
the one in the right graph is non-normalizable.
}
\label{fig:old_example_1}
\end{center}
\end{figure}

The analysis for the $U(1)$ part goes almost parallel to those above. 
Firstly, we have from Eq.~(\ref{eq:betas})
\be
\beta_1(r) = v^2\sqrt{\frac{2}{5(1+r_3^2)^4}+\frac{3}{5(1+r_2^2)^4}}.
\ee
This again leads to Schr\"odinger problems with the potentials
\be
\vcal_{\alpha=1} = \frac{\p_a^2\beta_1}{\beta_1},\quad
\tilde{\vcal}_{\alpha=1} = \frac{\p_a^2\beta_1^{-1}}{\beta_1^{-1}}.
\ee
Instead of showing form, we just display 
$\vcal_\alpha$ and its zero mode $\beta_\alpha$ in 
Fig.~\ref{fig:example_321}.
Reflecting the fact that $\beta_1$ is a weighted average of 
$|\beta_2|$ and $|\beta_3|$, the corresponding potential 
$\vcal_1$ has two attractive valleys around the vortices at 
$(x^4,x^5) = (\pm 2,0)$.
Hence the zero mode wave function of $Y_\mu$ is concentrated 
around both the vortex associated to $SU(3)$ and $SU(2)$. 
This results in an interesting difference between wave functions 
of $Y_\mu$ compared to $G_\mu$ and $W_\mu$. 
The spectra of the gluons and W bosons do not depend on 
the separation between vortices. The gluon (W boson) has a single massless 
mode localized around the vortex associated to $SU(3)$ ($SU(2)$).
On the other hand, the spectrum of $Y_\mu$ depends on the vortex 
separations, though the single zero mode always exists. 
Suppose that the vortices associated to $SU(3)$ and $SU(2)$ are 
infinitely separated, then two potential wells are also 
separated infinitely. 
Each of the isolated well has a localized 
zero mode. Hence the whole spectrum should include two massless 
modes for $U(1)$ gauge field.
If we, however, bring the vortices again together the degenerate zero modes now split (level repulsion) by a quantum tunneling effect between the two wells.
The lowest one remains massless while the other one is lifted 
by an exponentially suppressed nonperturbative tunneling effect.
Note also that if the vortices associated to $SU(3)$ and $SU(2)$ 
are completely coincident, the $SU(5)$ is unbroken and $\beta_1$ 
becomes identical to $\beta_3$ and $\beta_2$. As we saw above, 
$\beta_3$ ($\beta_3$) admits only a single discrete spectrum 
which is massless, so is $\beta_1$ at the coincident limit.
Therefore, the lifted zero mode that is normalizable for 
well-separated 3-2 splitting background should enter into 
continuum spectrum at a particular value of the separation. 
Concrete examples are given in Fig.~\ref{fig:old_example_1}. 
Hence, an extra massive mode can exist in $Y_\mu$ 
unlike in $G_\mu$ and $W_\mu$. This extra massive vector particle 
can be an evidence for the underlying background of vortices. 
It may be observable if the 3-2 split configuration with a large 
separation of the vortices is realized as the stabilized 
configuration.

\begin{figure}[h]
\begin{center}
\includegraphics[width=13cm]{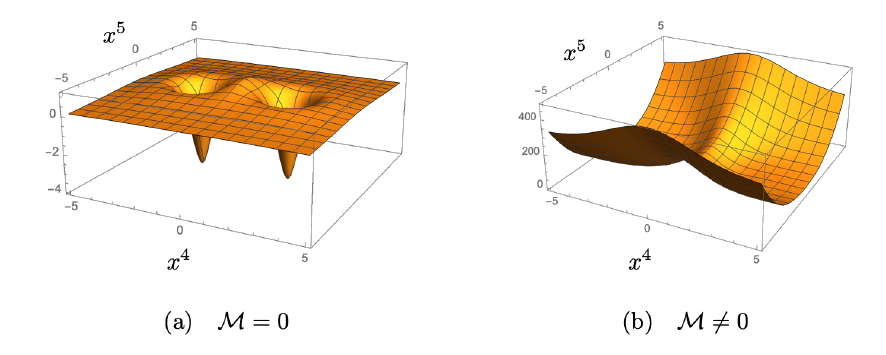}
\caption{The potential $V_X = \frac{\p_a^2 \beta_X}{\beta_X} 
+ \mcal^2$. The panel (a) shows $V_X$ without $\mcal^2$
and (b) with $\mcal^2$.}
\label{fig:Vx}
\end{center}
\end{figure}

To see mass spectra of the broken sector, we next need the other 
$\beta$'s from Eq.~(\ref{eq:betas})
\be
\beta_X = v^2\sqrt{\frac{1}{2(1+r_3^2)^4}+\frac{1}{2(1+r_2^2)^4}},\quad
\beta_\phi = v\left(\frac{r_3e^{i\theta_3}}{\sqrt{1+r_3^2}} 
- \frac{r_2e^{i\theta_2}}{\sqrt{1+r_2^2}}\right).
\ee
We can compute the Laplacian $\bm{\Delta}$ for the four-dimensional 
component $X_\mu$
\be
\bm{\Delta} = \triangle^X + \mcal^2 = \sum_{a=4,5}\left(-\p_a^2 
+ \frac{\p_a^2 \beta_X}{\beta_X}\right) + \mcal^2,\quad
\mcal^2 = \frac{|\beta_\phi|^2}{\beta_X^2}.
\ee
This is also a bit complicated, so we do not show the explicit expression.
If we omit $\mcal$, $\beta_X$ is quite similar to $\beta_1$ except 
for the weights for taking average.
Then, there exists a single normalizable zero mode which is 
proportional to $\beta_X$. However, this zero mode
is completely swept out by the additional term $\mcal^2$ which 
dominates the potential. 
Fig.~\ref{fig:Vx} shows a concrete example of the potential with and without $\mcal^2$. 
The vortex potential wells are clearly visible for $\mcal^2=0$, 
but it is almost washed out for $\mcal^2\not=0$. 
Although we do not find eigenvalues and eigenfunctions exactly, 
we are sure that there are no massless modes and a mass gap of 
order GUT scale exists when $\mcal^2$ is of the order of GUT scale. 
Showing the absence of the additional massless vector field 
rigorously is an important problem from the phenomenological viewpoint 
and it is left as a future work. 

\section{Conclusions and discussion}

In this work we examined issues associated to the SM gauge fields, 
gluons, weak bosons, and hypercharge $U(1)$ gauge fields specifically their localization on the non-Abelian vortices as 3-branes in six-dimensional 
spacetime.
Our six-dimensional Lagrangian has $SU(5)$ gauge symmetry, so the 
model has an aspect of $SU(5)$ GUT in addition to a fat brane-world 
scenario. 
The model also have additional global $U(1)$ symmetry, 
which is broken spontaneously in the bulk and gives rise to 
topologically stable vortices of the non-Abelian kind. 
On the other hand, the $SU(5)$ gauge symmetry is unbroken in 
the bulk, but breaks only {\it locally} near the cores of vortices. 
Hence, to which subgroup the $SU(5)$ gauge symmetry breaks depends 
on how many vortices are generated and where they are located.

In Sec.~\ref{sec:model} we studied two specific models which 
exhibit the desired symmetry breaking $SU(5) \to SU(3)\times SU(2) 
\times U(1)$ by the non-Abelian vortices.
The first model in Sec.~\ref{sec:simplest_non-Abelian_vortex} 
admits the embedding of usual $U(1)$ global vortices into diagonal 
elements of the 5 by 5 matrix field $T$. 
The embedded vortex is a 
$\frac{1}{5}$ fractionally quantized non-Abelian vortex.
There are five different species of the fractional non-Abelian 
vortex associated with each diagonal element of $T$.
The desired configuration can be constructed by putting a vortex 
with a common vorticity $k_3$ in the first three entries and 
another vortex with a common vorticity $k_2$ in the remaining two 
entries of the diagonal elements of $T$ given as $(k_3,k_3,k_3,k_2,k_2)$. 
However, the position (moduli) of vortices can be freely changed 
in this first model, leading to unwanted patterns of symmetry 
breaking such $SU(5)\to U(1)^4$. 
We need to remove the zero modes of such a deformation. 
This moduli stabilization problem was solved in the second 
model by adding simple deformation potentials in Sec.~\ref{sec:improve}.
The key point is to deal with the traceless part $\hat T$ (adjoint 
representation) of $T$ separately from the trace part $T_0$, since 
the vortex separation corresponds to nonvanishing adjoint fields.
Our simple potential induces a domain wall attached to the 
fractional non-Abelian vortex. 
The domain walls glue the fractional non-Abelian vortices, and 
we found that the fractional non-Abelian vortices are confined 
to form $SU(5)$ singlet. 
This dynamical process aligns the vorticity, so that a stable 
configuration has to have identical vorticities 
in all diagonal entries of $T$. Thus we found that the 
domain walls tend to fully confine the fractional non-Abelian vortices, 
but this is not our desired solution since the $SU(5)$ gauge 
symmetry is unbroken. 
This point is resolved by a short range repulsive interaction 
between the vortices. 
Competition between the long range attraction by the domain wall 
and the short range repulsion results in a singlet combination clustering into a 3-2 splitting 
molecule. 
Namely, the five vortices with identical vorticity split into 
a molecule configuration with three vortices at a point 
(associated to $SU(3)$) and two vortices (associated to $SU(2)$) 
at another point separated by a small distance. 
This is phenomenologically important once fermions are introduced 
though we did not explicitly deal with them in this paper.
The fact that there are identical vorticities associated to $SU(3)$ and 
to $SU(2)$ ensures the same number of fermion zero modes 
in the fundamental representation of $SU(3)$ (quark) and of $SU(2)$ 
(lepton) are localized. 

In Sec.~\ref{sec:fluctation} we turned to clarify the localization 
of the SM gauge fields on a 3-2 splitting background. 
We performed a standard fluctuation study by introducing small 
fluctuations of all the fields and expanded the Lagrangian to 
the quadratic order of fluctuations. 
We can then extract physical informations such as mass spectra 
of fluctuations. 
The procedure has a number of complications: 
\begin{enumerate}
\item 
The background configuration is inhomogeneous vortex background, 
which is not axially symmetric because of the 3-2 splitting molecule. 
\item
The $SU(5)$ gauge symmetry is locally broken to 
$SU(3) \times SU(2) \times U(1)$, and we need to treat 
the gauge fields in the unbroken and broken sectors differently.
\item
We need a gauge fixing suitable for the vortex molecule background.
\item
We need to treat four-dimensional and extra-dimensional components 
of the gauge fields differently. 
\end{enumerate}
We succeeded in showing that the massless gauge fields corresponding 
to the SM gauge group are localized on the non-Abelian vortex molecule, 
thanks to a field-dependent gauge kinetic function $\bcal\bcal^\dag$ 
in Eq.~(\ref{eq:lag}). 
In order to perform the fluctuation analysis efficiently, we 
developed a vector-analysis like method using derivative operators 
$\vec D$ of the two extra dimensions ($x^4,x^5$) in Eq.~(\ref{eq:D_unbroken}). 
The method helps not only to make the expressions compact, but also 
to distinguish physical and unphysical modes. 
The Helmholtz-like theorem turned out to be important to decompose 
the extra-dimensional gauge field $\vec A$ of the unbroken sector 
into the rotation-free and divergence-free parts.
We found that the rotation-free part of $\vec A$ ($b$'s in 
Eq.~(\ref{eq:Lag_unbroken})) is unphysical, playing the role 
of NG bosons for the KK towers (except for the bottom) of the 
four-dimensional gauge fields $A_\mu$ (the SM gauge fields: 
gluons, W bosons, and hypercharge gauge fields). 
On the other hand, the divergence-free part of $\vec A$
($c$'s in Eq.~(\ref{eq:Lag_unbroken})) is physical but massless modes are absent.
In short, the vortex effective theory contains the SM gauge fields 
as the only massless modes, and an infinite towers of 
massive KK modes. 
The extra-dimensional components ($A_4$ and $A_5$) provide only one 
(five-dimensional) scalar field degree of freedom whose spectrum 
does not have massless modes.
We applied the same kind of technique to the unbroken sector, 
although the analysis becomes more complicated because of the 
mixing with $T$ field. 
Although analogy to the three-dimensional vector analysis is not 
as complete as in the unbroken sector, 
we found the vector-analysis like method is still useful.
Similarly to the unbroken sector, the ``rotation-free'' component 
of $\bm{X}$ is found to be unphysical because it is absorbed by $X_\mu$.
The important difference from the unbroken sector is that {\it all} 
the KK modes of $X_\mu$ become massive, so that there are no massless 
vector modes in the unbroken sector.
The remaining three degrees of freedom, the divergence-free 
components, of $\bm{X}$ were only partially understood.
We gave the formal expression of the 4 by 4 Laplacian 
$\tilde{\bm \Delta}_{\rm v}$ in Eq.~(\ref{eq:tildeDelta_v}), whose 
eigenvalues depend on details of the model. 
We also demonstrated the low lying mass spectrum in a concrete 
model and pointed out the possibility that an exotic heavy (but 
not too heavy) vector field in the KK tower of the $U(1)$ hypercharge 
vector field may be observed.

Thus we provided a class of new models suitable for the gauge 
sector of the fat brane-world scenario with GUT in six spacetime 
dimensions by the non-Abelian vortices. 
As a future work, firstly, we need to include fermions to 
complete our mission. As usual in $SU(5)$ models, 
it is natural to have $\bm{5}$ and $\bm{10}$ representations 
of $SU(5)$ as fermions. 
The most important advantage of using the vortices as the host 
3-branes is that the six-dimensional theory gives the same number 
of fermion zero modes for the fundamental representation of 
$SU(3)$ (quarks) and $SU(2)$ (leptons) forming generations. 
Namely, the number of the fermion generations is identical to 
vorticity (the topological number) of the background solution.  
Secondly, we also need to include a Higgs field which breaks 
the electroweak symmetry. 
There is a longstanding issue known as the double-triplet splitting 
problem in $SU(5)$ GUT models. 
Whether the 3-2 split non-Abelian vortices configuration can 
account for the doublet-triplet problem without fine tunings 
is a challenging future problem. 
Other phenomenologically challenging issues include hierarchy in 
quark/lepton masses and mixing matrices, possible right-handed neutrinos, 
and proton decay.


\section*{Acknowledgements}
This work is supported in part by JSPS Grant-in-Aid for Scientific  
Research KAKENHI Grant No. JP21K03565 (M. A.), JP19K03839 (M. E.) and 18H01217 (N. S.).
The work of M. E. is supported in part by MEXT KAKENHI Grant-in-Aid for Scientific 
Research on Innovative Areas “Discrete Geometric Analysis for Materials Design”  
No.JP17H06462 from the MEXT of Japan. 
F. Blaschke would like to express his acknowledgment for the institutional support of the Research Centre for Theoretical Physics and Astrophysics, Institute of Physics, Silesian University in Opava.

\begin{appendix}
\section{Generalized Helmholtz's decomposition}
\label{sec:app1}

Here we give a proof that a vector field $\vec A$ can be always decomposed
into the rotation- and divergence-free components as Eq.~(\ref{eq:Helmholtz}).
Firstly, we define a projection operator which projects out $\vec A$ onto the rotation-free component
\be
P = \vec D \circ \left(\vec D^\dagger \cdot \vec D \circ\right)^{-1}\vec D^\dagger \cdot 
= \, {\rm grad} \, \triangle^{-1}\, {\rm div}\,.
\ee
This is a projection operator since it satisfies $P^2 = P$ as
\be
P^2 = \left(\vec D \circ \left(\vec D^\dagger \cdot \vec D \circ\right)^{-1}\vec D^\dagger \cdot \right)^2 = P.
\ee
Therefore, we can always decompose the vector $\vec A = P\vec A + (1-P)\vec A$.
However, $P$ is well-defined only when the inverse Laplacian $\triangle^{-1}$ is well-defined.
A dangerous case occurs if $\triangle^{-1}$ acts on its zero mode $\triangle B_0 = 0$ (${\rm grad}\, B_0 = 0$) 
as is given in Eq.~(\ref{eq:mode_of_triangle}).
However, this is not the case for our case because $\triangle^{-1}$ acts right after the divergence ${\rm div}$.
Let $\vec E$ as an arbitrary two-component vector, and take an inner product as
\be
\left(B_0\,,\, {\rm div}\,\vec E\right) 
= \left({\rm grad}\,B_0, \vec E\right) 
= 0,
\ee
where we defined the inner product by 
$\left(A\,,\, B\right)  = \int dx^4dx^5\, A^*\, B$
and $\left(\vec A\,,\, \vec B\right)  = \int dx^4dx^5\, \vec A^\dag\, \vec B$.
Hence, we conclude that ${\rm div}\, \vec E$ does not include $B_0$, and therefore
$\triangle^{-1}\,{\rm div}$ is always well-defined, so is $P$.

Next, we show $P\vec A$ is rotation-free. This is trivial by definition because
\be
P \vec A = \, {\rm grad} \, \left(\triangle^{-1}\, {\rm div}\, \vec A\right),
\ee
and from Eq.~(\ref{eq:rot_grad}) we always have ${\rm rot}_{\rm v}\, P\vec A = 0$.

Let us next treat $(1-P)\vec A$. We expand it by eigenfunctions of the non-negative definite Hermitian operator 
$\tilde{\triangle}_{\rm v} = \, {\rm rot}_{\rm s}\,{\rm rot}_{\rm v}\,$
\be
\tilde{\triangle}_{\rm v} \vec C_n &=& \, {\rm rot}_{\rm s}\,{\rm rot}_{\rm v}\, \vec C_n = \tilde m_n^2 \vec C_n,\\
(1-P)\vec A &=& \sum_{n \ge 0} d_n \vec C_n.
\ee
Note that index $n$ of the eigenvalue starts at $n=0$ which corresponds to the lowest eigenvalue.
Indeed, the lowest eigenvalue is $\tilde m_0 = 0$ and the corresponding eigenfunction is 
\be
{\rm rot}_{\rm v}\, \vec C_0 = 0 \quad \to \quad \vec C_0 \propto \,{\rm grad}\,\xi,
\ee
with $\xi$ being an arbitrary function. 
We now show $d_0$ is zero. This is because
\be
\left((1-P)\vec A, \vec C_0\right) &\propto&
\left((1-P)\vec A, \,{\rm grad}\,\xi\right) \nonumber\\
&=& \left({\rm div}\,(1-P)\vec A, \xi\right) \nonumber\\
&=& \left({\rm div}\,\vec A -{\rm div}\,{\rm grad} \, \triangle^{-1}\, {\rm div}\, \vec A,\xi\right) = 0.
\ee
Therefore, the above expansion is modified as
\be
(1-P)\vec A = \sum_{n > 0} d_n \vec C_n.
\ee
Note also that all the positive eigenvalues $\tilde m_n^2$ ($n>0$) of $\tilde{\triangle}_{\rm v}$ correspond  one-to-one to those of 
the $\tilde{\triangle}_{\rm s}$ defined in Eq.~(\ref{eq:mode_of_triangle}). Actual correspondence is given by
\be
\vec C_n \propto {\rm rot}_{\rm s}\, C_n,\quad
C_n \propto {\rm rot}_{\rm v}\,\vec C_n.
\ee
This can be verified as follows for $n > 0$ as
\be
\tilde{\triangle}_{\rm v} \left({\rm rot}_{\rm s}\, C_n\right) 
= (\,{\rm rot}_{\rm s}\,{\rm rot}_{\rm v}\,) \,{\rm rot}_{\rm s}\, C_n
= \,{\rm rot}_{\rm s}\,\tilde{\triangle}_{\rm s} C_n = \tilde m_n^2 (\,{\rm rot}_{\rm s}\,C_n).
\ee
It is straightforward to show the other equation.
Hence, we now arrive at the desired expression
\be
(1-P)\vec A = \, {\rm rot}_{\rm s}\,\left(\sum_{n > 0} d_n' \vec C_n\right),
\ee
that implies $(1-P)A$ is divergence free.

\end{appendix}

\bibliographystyle{jhep}


\begin{thebibliography}{99}

\bibitem{ArkaniHamed:1998rs}
N.~Arkani-Hamed, S.~Dimopoulos and G.~R.~Dvali,
``The Hierarchy problem and new dimensions at a millimeter,''
Phys. Lett. B \textbf{429}, 263-272 (1998)
doi:10.1016/S0370-2693(98)00466-3
[arXiv:hep-ph/9803315 [hep-ph]].

\bibitem{Antoniadis:1998ig}
I.~Antoniadis, N.~Arkani-Hamed, S.~Dimopoulos and G.~R.~Dvali,
``New dimensions at a millimeter to a Fermi and superstrings at a TeV,''
Phys. Lett. B \textbf{436}, 257-263 (1998)
doi:10.1016/S0370-2693(98)00860-0
[arXiv:hep-ph/9804398 [hep-ph]].

\bibitem{Randall:1999ee}
L.~Randall and R.~Sundrum,
``A Large mass hierarchy from a small extra dimension,''
Phys. Rev. Lett. \textbf{83}, 3370-3373 (1999)
doi:10.1103/PhysRevLett.83.3370
[arXiv:hep-ph/9905221 [hep-ph]].

\bibitem{Dvali:1996bg}
G.~R.~Dvali and M.~A.~Shifman,
``Dynamical compactification as a mechanism of spontaneous 
supersymmetry breaking,''
Nucl. Phys. B \textbf{504}, 127-146 (1997)
doi:10.1016/S0550-3213(97)00420-3
[arXiv:hep-th/9611213 [hep-th]].

\bibitem{Jackiw:1975fn} 
  R.~Jackiw and C.~Rebbi,
  ``Solitons with Fermion Number 1/2,''
  Phys.\ Rev.\ D {\bf 13}, 3398 (1976).
  doi:10.1103/PhysRevD.13.3398

\bibitem{Rubakov:1983bb} 
  V.~A.~Rubakov and M.~E.~Shaposhnikov,
  ``Do We Live Inside a Domain Wall?,''
  Phys.\ Lett.\  {\bf 125B}, 136 (1983).
  doi:10.1016/0370-2693(83)91253-4.

\bibitem{ArkaniHamed:1999dc}
N.~Arkani-Hamed and M.~Schmaltz,
``Hierarchies without symmetries from extra dimensions,''
Phys. Rev. D \textbf{61}, 033005 (2000)
doi:10.1103/PhysRevD.61.033005
[arXiv:hep-ph/9903417 [hep-ph]].


\bibitem{Dvali:1996xe}
G.~R.~Dvali and M.~A.~Shifman,
``Domain walls in strongly coupled theories,''
Phys. Lett. B \textbf{396}, 64-69 (1997)
[erratum: Phys. Lett. B \textbf{407}, 452 (1997)]
doi:10.1016/S0370-2693(97)00131-7
[arXiv:hep-th/9612128 [hep-th]].

  

\bibitem{Maru:2003mx}
  N.~Maru and N.~Sakai,
  ``Localized gauge multiplet on a wall,''
  Prog.\ Theor.\ Phys.\  {\bf 111} (2004) 907
  [arXiv:hep-th/0305222].



\bibitem{Ohta:2010fu}
K.~Ohta and N.~Sakai,
``Non-Abelian Gauge Field Localized on Walls with Four-Dimensional World Volume,''
Prog. Theor. Phys. \textbf{124}, 71-93 (2010)
[erratum: Prog. Theor. Phys. \textbf{127}, 1133 (2012)]
doi:10.1143/PTP.124.71
[arXiv:1004.4078 [hep-th]].

\bibitem{Arai:2012cx}
M.~Arai, F.~Blaschke, M.~Eto and N.~Sakai,
``Matter Fields and Non-Abelian Gauge Fields Localized on Walls,''
PTEP \textbf{2013}, 013B05 (2013)
doi:10.1093/ptep/pts050
[arXiv:1208.6219 [hep-th]].

\bibitem{Arai:2013mwa}
M.~Arai, F.~Blaschke, M.~Eto and N.~Sakai,
``Stabilizing matter and gauge fields localized on walls,''
PTEP \textbf{2013}, no.9, 093B01 (2013)
doi:10.1093/ptep/ptt064
[arXiv:1303.5212 [hep-th]].

\bibitem{Arai:2017lfv}
M.~Arai, F.~Blaschke, M.~Eto and N.~Sakai,
Phys. Rev. D \textbf{96}, no.11, 115033 (2017)
doi:10.1103/PhysRevD.96.115033
[arXiv:1703.00351 [hep-th]].

\bibitem{Arai:2017ntb}
M.~Arai, F.~Blaschke, M.~Eto and N.~Sakai,
``Non-Abelian Gauge Field Localization on Walls and Geometric 
Higgs Mechanism,''
PTEP \textbf{2017}, no.5, 053B01 (2017)
doi:10.1093/ptep/ptx047
[arXiv:1703.00427 [hep-th]].

\bibitem{Arai:2018uoy}
M.~Arai, F.~Blaschke, M.~Eto and N.~Sakai,
``Localization of the Standard Model via the Higgs mechanism and a finite electroweak monopole from non-compact five dimensions,''
PTEP \textbf{2018}, no.8, 083B04 (2018)
doi:10.1093/ptep/pty083
[arXiv:1802.06649 [hep-ph]].

\bibitem{Arai:2018rwf}
M.~Arai, F.~Blaschke, M.~Eto and N.~Sakai,
``Localized non-Abelian gauge fields in non-compact extra-dimensions,''
PTEP \textbf{2018}, no.6, 063B02 (2018)
doi:10.1093/ptep/pty057
[arXiv:1801.02498 [hep-th]].

\bibitem{Eto:2019weg}
M.~Eto and M.~Kawaguchi,
``Localization of gauge bosons and the Higgs mechanism on topological solitons in higher dimensions,''
JHEP \textbf{10}, 098 (2019)
doi:10.1007/JHEP10(2019)098
[arXiv:1907.04573 [hep-th]].

\bibitem{Arai:2018hao}
M.~Arai, F.~Blaschke, M.~Eto and N.~Sakai,
``Massless bosons on domain walls: Jackiw-Rebbi-like mechanism for bosonic fields,''
Phys. Rev. D \textbf{100}, no.9, 095014 (2019)
doi:10.1103/PhysRevD.100.095014
[arXiv:1811.08708 [hep-th]].


\bibitem{Oda:2000zc}
I.~Oda,
``Localization of matters on a string - like defect,''
Phys. Lett. B \textbf{496} (2000), 113-121
doi:10.1016/S0370-2693(00)01284-3
[arXiv:hep-th/0006203 [hep-th]].



\bibitem{Dvali:2000rx}
  G.~R.~Dvali, G.~Gabadadze and M.~A.~Shifman,
  ``(Quasi)localized gauge field on a brane: Dissipating cosmic radiation to extra dimensions?,''
  Phys.\ Lett.\ B {\bf 497}, 271 (2001)
  doi:10.1016/S0370-2693(00)01329-0
  [hep-th/0010071].
 
 \bibitem{Kehagias:2000au}
  A.~Kehagias and K.~Tamvakis,
  ``Localized gravitons, gauge bosons and chiral fermions in smooth spaces generated by a bounce,''
  Phys.\ Lett.\ B {\bf 504}, 38 (2001)
 doi:10.1016/S0370-2693(01)00274-X
 [hep-th/0010112].

\bibitem{Oda:2000dd}
I.~Oda,
``Localization of bulk fields on AdS(4) brane in AdS(5),''
Phys. Lett. B \textbf{508} (2001), 96-102
doi:10.1016/S0370-2693(01)00376-8
[arXiv:hep-th/0012013 [hep-th]].

\bibitem{Oda:2001ux}
I.~Oda,
``A new mechanism for trapping of photon,''
[arXiv:hep-th/0103052 [hep-th]].


\bibitem{Oda:2001yx}
I.~Oda,
``Trapping of nonAbelian gauge fields on a brane,''
[arXiv:hep-th/0103257 [hep-th]].
  
\bibitem{Dubovsky:2001pe}
  S.~L.~Dubovsky and V.~A.~Rubakov,
  ``On models of gauge field localization on a brane,''
  Int.\ J.\ Mod.\ Phys.\ A {\bf 16}, 4331 (2001)
  doi:10.1142/S0217751X01005286
  [hep-th/0105243].


\bibitem{Ghoroku:2001zu}
  K.~Ghoroku and A.~Nakamura,
  ``Massive vector trapping as a gauge boson on a brane,''
  Phys.\ Rev.\ D {\bf 65}, 084017 (2002) 
 doi:10.1103/PhysRevD.65.084017
 [hep-th/0106145].
   
  \bibitem{Akhmedov:2001ny}
  E.~K.~Akhmedov,
  ``Dynamical localization of gauge fields on a brane,''
  Phys.\ Lett.\ B {\bf 521}, 79 (2001)
  doi:10.1016/S0370-2693(01)01176-5
  [hep-th/0107223].

\bibitem{Kogan:2001wp}
  I.~I.~Kogan, S.~Mouslopoulos, A.~Papazoglou and G.~G.~Ross,
  ``Multilocalization in multibrane worlds,''
  Nucl.\ Phys.\ B {\bf 615}, 191 (2001)
  doi:10.1016/S0550-3213(01)00424-2
 [hep-ph/0107307].
   
 \bibitem{Abe:2002rj}
  H.~Abe, T.~Kobayashi, N.~Maru and K.~Yoshioka,
  ``Field localization in warped gauge theories,''
  Phys.\ Rev.\ D {\bf 67}, 045019 (2003)  
  doi:10.1103/PhysRevD.67.045019
 [hep-ph/0205344].

\bibitem{Laine:2002rh}
  M.~Laine, H.~B.~Meyer, K.~Rummukainen and M.~Shaposhnikov,
  ``Localization and mass generation for nonAbelian gauge fields,''
  JHEP {\bf 0301}, 068 (2003)
  doi:10.1088/1126-6708/2003/01/068
  [hep-ph/0211149].
  
\bibitem{Batell:2006dp}
  B.~Batell and T.~Gherghetta,
  ``Yang-Mills Localization in Warped Space,''
  Phys.\ Rev.\ D {\bf 75}, 025022 (2007) 
 doi:10.1103/PhysRevD.75.025022
 [hep-th/0611305].

\bibitem{Guerrero:2009ac}
  R.~Guerrero, A.~Melfo, N.~Pantoja and R.~O.~Rodriguez,
  ``Gauge field localization on brane worlds,''
  Phys.\ Rev.\ D {\bf 81}, 086004 (2010) 
  doi:10.1103/PhysRevD.81.086004
 [arXiv:0912.0463 [hep-th]].
  
\bibitem{Cruz:2010zz}
  W.~T.~Cruz, M.~O.~Tahim and C.~A.~S.~Almeida,
  ``Gauge field localization on a dilatonic deformed brane,''
  Phys.\ Lett.\ B {\bf 686}, 259 (2010).
  doi:10.1016/j.physletb.2010.02.064

\bibitem{Chumbes:2011zt}
  A.~E.~R.~Chumbes, J.~M.~Hoff da Silva and M.~B.~Hott,
  ``A model to localize gauge and tensor fields on thick branes,''
  Phys.\ Rev.\ D {\bf 85}, 085003 (2012) 
 doi:10.1103/PhysRevD.85.085003
 [arXiv:1108.3821 [hep-th]].
    
\bibitem{Germani:2011cv}
  C.~Germani,
  ``Spontaneous localization on a brane via a gravitational mechanism,''
  Phys.\ Rev.\ D {\bf 85}, 055025 (2012)
  doi:10.1103/PhysRevD.85.055025
  [arXiv:1109.3718 [hep-ph]].

\bibitem{Delsate:2011aa}
  T.~Delsate and N.~Sawado,
  ``Localizing modes of massive fermions and a U(1) gauge field in the inflating baby-skyrmion branes,''
  Phys.\ Rev.\ D {\bf 85}, 065025 (2012) 
  doi:10.1103/PhysRevD.85.065025
  [arXiv:1112.2714 [gr-qc]].
  
\bibitem{Cruz:2012kd}
  W.~T.~Cruz, A.~R.~P.~Lima and C.~A.~S.~Almeida,
  Phys.\ Rev.\ D {\bf 87}, no. 4, 045018 (2013)
  doi:10.1103/PhysRevD.87.045018
  [arXiv:1211.7355 [hep-th]].

\bibitem{Herrera-Aguilar:2014oua}
  A.~Herrera-Aguilar, A.~D.~Rojas and E.~Santos-Rodriguez,
  ``Localization of gauge fields in a tachyonic de Sitter thick braneworld,''
  Eur.\ Phys.\ J.\ C {\bf 74}, no. 4, 2770 (2014)
  doi:10.1140/epjc/s10052-014-2770-1
  [arXiv:1401.0999 [hep-th]].
  
\bibitem{Zhao:2014gka}
  Z.~H.~Zhao, Y.~X.~Liu and Y.~Zhong,
  ``U(1) gauge field localization on a Bloch brane with Chumbes-Holf da Silva-Hott mechanism,''
  Phys.\ Rev.\ D {\bf 90}, no. 4, 045031 (2014)
  doi:10.1103/PhysRevD.90.045031
  [arXiv:1402.6480 [hep-th]].

\bibitem{Vaquera-Araujo:2014tia}
  C.~A.~Vaquera-Araujo and O.~Corradini,
  ``Localization of abelian gauge fields on thick branes,''
  Eur.\ Phys.\ J.\ C {\bf 75}, no. 2, 48 (2015)
  doi:10.1140/epjc/s10052-014-3251-2
  [arXiv:1406.2892 [hep-th]].

\bibitem{Alencar:2014moa} 
  G.~Alencar, R.~R.~Landim, M.~O.~Tahim and R.~N.~Costa Filho,
  ``Gauge Field Localization on the Brane Through Geometrical Coupling,''
  Phys.\ Lett.\ B {\bf 739}, 125 (2014)
  doi:10.1016/j.physletb.2014.10.040
  [arXiv:1409.4396 [hep-th]].
  

\bibitem{Davies:2007xr}
R.~Davies, D.~P.~George and R.~R.~Volkas,
``The Standard model on a domain-wall brane,''
Phys. Rev. D \textbf{77}, 124038 (2008)
doi:10.1103/PhysRevD.77.124038
[arXiv:0705.1584 [hep-ph]].

\bibitem{Davidson:2007cf}
A.~Davidson, D.~P.~George, A.~Kobakhidze, R.~R.~Volkas and K.~C.~Wali,
``SU(5) grand unification on a domain-wall brane from an E(6)-invariant action,''
Phys. Rev. D \textbf{77}, 085031 (2008)
doi:10.1103/PhysRevD.77.085031
[arXiv:0710.3432 [hep-ph]].

\bibitem{Thompson:2009uk}
J.~E.~Thompson and R.~R.~Volkas,
``SO(10) domain-wall brane models,''
Phys. Rev. D \textbf{80}, 125016 (2009)
doi:10.1103/PhysRevD.80.125016
[arXiv:0908.4122 [hep-ph]].

\bibitem{Callen:2010mx}
B.~D.~Callen and R.~R.~Volkas,
``Fermion masses and mixing in a 4+1-dimensional SU(5) domain-wall brane model,''
Phys. Rev. D \textbf{83}, 056004 (2011)
doi:10.1103/PhysRevD.83.056004
[arXiv:1008.1855 [hep-ph]].

\bibitem{Callen:2012kd}
B.~D.~Callen and R.~R.~Volkas,
``Large lepton mixing angles from a 4+1-dimensional SU(5) x A(4) domain-wall braneworld model,''
Phys. Rev. D \textbf{86}, 056007 (2012)
doi:10.1103/PhysRevD.86.056007
[arXiv:1205.3617 [hep-ph]].

\bibitem{Okada:2017omx}
N.~Okada, D.~Raut and D.~Villalba,
``Domain-Wall Standard Model in non-compact 5D and LHC phenomenology,''
Mod. Phys. Lett. A \textbf{34}, no.10, 1950080 (2019)
doi:10.1142/S0217732319500809
[arXiv:1712.09323 [hep-ph]].

\bibitem{Okada:2019fgm}
N.~Okada, D.~Raut and D.~Villalba,
``Fermion Mass Hierarchy and Phenomenology in the 5D Domain Wall Standard Model,''
JHEP \textbf{10}, 259 (2019)
doi:10.1007/JHEP10(2019)259
[arXiv:1904.10308 [hep-ph]].

\bibitem{Libanov:2005mv}
M.~V.~Libanov and E.~Y.~Nugaev,
``Properties of the Higgs particle in a model involving a single unified fermion generation,''
Phys. Atom. Nucl. \textbf{70}, 864-870 (2007)
doi:10.1134/S1063778807050092
[arXiv:hep-ph/0512223 [hep-ph]].

\bibitem{Frere:2000dc}
J.~M.~Frere, M.~V.~Libanov and S.~V.~Troitsky,
``Three generations on a local vortex in extra dimensions,''
Phys. Lett. B \textbf{512}, 169-173 (2001)
doi:10.1016/S0370-2693(01)00696-7
[arXiv:hep-ph/0012306 [hep-ph]].


\bibitem{Eto:2004ii}
M.~Eto, M.~Nitta and N.~Sakai,
``Effective theory on non-Abelian vortices in six dimensions,''
Nucl. Phys. B \textbf{701}, 247-272 (2004)
doi:10.1016/j.nuclphysb.2004.09.003
[arXiv:hep-th/0405161 [hep-th]].

\bibitem{Balachandran:2002je}
A.~P.~Balachandran and S.~Digal,
``NonAbelian topological strings and metastable states in linear sigma model,''
Phys. Rev. D \textbf{66}, 034018 (2002)
doi:10.1103/PhysRevD.66.034018
[arXiv:hep-ph/0204262 [hep-ph]].


\bibitem{Nitta:2007dp}
M.~Nitta and N.~Shiiki,
``Non-Abelian Global Strings at Chiral Phase Transition,''
Phys. Lett. B \textbf{658}, 143-147 (2008)
doi:10.1016/j.physletb.2007.10.055
[arXiv:0708.4091 [hep-ph]].


\bibitem{Nakano:2007dq}
E.~Nakano, M.~Nitta and T.~Matsuura,
``Interactions of non-Abelian global strings,''
Phys. Lett. B \textbf{672}, 61-64 (2009)
doi:10.1016/j.physletb.2008.11.049
[arXiv:0708.4092 [hep-ph]].


\bibitem{Eto:2009wu}
M.~Eto, E.~Nakano and M.~Nitta,
``Non-Abelian Global Vortices,''
Nucl. Phys. B \textbf{821}, 129-150 (2009)
doi:10.1016/j.nuclphysb.2009.06.013
[arXiv:0903.1528 [hep-ph]].

\bibitem{Eto:2013bxa}
M.~Eto, Y.~Hirono and M.~Nitta,
``Domain Walls and Vortices in Chiral Symmetry Breaking,''
PTEP \textbf{2014}, no.3, 033B01 (2014)
doi:10.1093/ptep/ptu013
[arXiv:1309.4559 [hep-ph]].





\bibitem{Son:2001td}
D.~T.~Son and M.~A.~Stephanov,
``Domain walls in two-component Bose-Einstein condensates,''
Phys. Rev. A \textbf{65}, 063621 (2002)
doi:10.1103/PhysRevA.65.063621
[arXiv:cond-mat/0103451 [cond-mat.soft]].

\bibitem{Kasamatsu:2004tvg}
K.~Kasamatsu, M.~Tsubota and M.~Ueda,
``Vortex molecules in coherently coupled two-component Bose-Einstein condensates,''
Phys. Rev. Lett. \textbf{93}, no.25, 250406 (2004)
doi:10.1103/PhysRevLett.93.250406
[arXiv:cond-mat/0406150 [cond-mat.mes-hall]].

\bibitem{Eto:2011wp}
M.~Eto, K.~Kasamatsu, M.~Nitta, H.~Takeuchi and M.~Tsubota,
``Interaction of half-quantized vortices in two-component Bose-Einstein condensates,''
Phys. Rev. A \textbf{83}, 063603 (2011)
doi:10.1103/PhysRevA.83.063603
[arXiv:1103.6144 [cond-mat.quant-gas]].

\bibitem{Eto:2012rc}
M.~Eto and M.~Nitta,
``Vortex trimer in three-component Bose-Einstein condensates,''
Phys. Rev. A \textbf{85}, 053645 (2012)
doi:10.1103/PhysRevA.85.053645
[arXiv:1201.0343 [cond-mat.quant-gas]].

\bibitem{Eto:2013spa}
M.~Eto and M.~Nitta,
``Vortex graphs as N-omers and CP(N-1) Skyrmions in N-component Bose-Einstein condensates,''
EPL \textbf{103}, no.6, 60006 (2013)
doi:10.1209/0295-5075/103/60006
[arXiv:1303.6048 [cond-mat.quant-gas]].

\bibitem{Nitta:2013eaa}
M.~Nitta, M.~Eto and M.~Cipriani,
``Vortex molecules in Bose-Einstein condensates,''
J. Low Temp. Phys. \textbf{175}, 177-188 (2013)
doi:10.1007/s10909-013-0925-3
[arXiv:1307.4312 [cond-mat.quant-gas]].

\bibitem{Kasamatsu:2015cia}
K.~Kasamatsu, M.~Eto and M.~Nitta,
``Short-range intervortex interaction and interacting dynamics of half-quantized vortices in two-component Bose-Einstein condensates,''
Phys. Rev. A \textbf{93}, no.1, 013615 (2016)
doi:10.1103/PhysRevA.93.013615
[arXiv:1510.00139 [cond-mat.quant-gas]].

\bibitem{Tylutki:2016mgy}
M.~Tylutki, L.~P.~Pitaevskii, A.~Recati and S.~Stringari,
``Confinement and precession of vortex pairs in coherently coupled Bose-Einstein condensates,''
Phys. Rev. A \textbf{93}, no.4, 043623 (2016)
doi:10.1103/PhysRevA.93.043623
[arXiv:1601.03695 [cond-mat.quant-gas]].

\bibitem{Eto:2017rfr}
M.~Eto and M.~Nitta,
``Confinement of half-quantized vortices in coherently coupled Bose-Einstein condensates: Simulating quark confinement in a QCD-like theory,''
Phys. Rev. A \textbf{97}, no.2, 023613 (2018)
doi:10.1103/PhysRevA.97.023613
[arXiv:1702.04892 [cond-mat.quant-gas]].

\bibitem{Kobayashi:2018ezm}
M.~Kobayashi, M.~Eto and M.~Nitta,
``Berezinskii-Kosterlitz-Thouless Transition of Two-Component Bose Mixtures with Intercomponent Josephson Coupling,''
Phys. Rev. Lett. \textbf{123}, no.7, 075303 (2019)
doi:10.1103/PhysRevLett.123.075303
[arXiv:1802.08763 [cond-mat.stat-mech]].

\bibitem{Eto:2019uhe}
M.~Eto, K.~Ikeno and M.~Nitta,
``Collision dynamics and reactions of fractional vortex molecules in coherently coupled Bose-Einstein condensates,''
Phys. Rev. Res. \textbf{2}, no.3, 033373 (2020)
doi:10.1103/PhysRevResearch.2.033373
[arXiv:1912.09014 [cond-mat.quant-gas]].

\bibitem{Yang:2020wes}
W.~C.~Yang, C.~Y.~Xia, M.~Nitta and H.~B.~Zeng,
``Fractional and Integer Vortex Dynamics in Strongly Coupled Two-component Bose-Einstein Condensates from AdS/CFT Correspondence,''
Phys. Rev. D \textbf{102}, no.4, 046012 (2020)
doi:10.1103/PhysRevD.102.046012
[arXiv:2003.09423 [cond-mat.quant-gas]].



\bibitem{Eto:2021nle}
M.~Eto and M.~Nitta,
``Minimum non-Abelian vortices and their confinement in three flavor dense QCD,''
[arXiv:2103.13011 [hep-ph]].

\end{thebibliography}
\end{document}